\documentclass[acmsmall,manuscript,screen]{acmart}

\AtBeginDocument{%
  \providecommand\BibTeX{{%
    \normalfont B\kern-0.5em{\scshape i\kern-0.25em b}\kern-0.8em\TeX}}}

\setcopyright{acmcopyright}
\copyrightyear{2021}
\acmYear{2021}
\acmDOI{to be assigned}

 \acmJournal{TOMS}
\acmPrice{15.00}
\acmISBN{978-1-4503-XXXX-X/18/06}

\usepackage{tikz}
\usetikzlibrary{snakes}
\usepackage{pgfplots}
\pgfplotsset{compat=newest}
\usepackage{pgfplotstable} % For \pgfplotstableread
\usepgfplotslibrary{patchplots}
\pgfplotscreateplotcyclelist{cbw}{%
    ultra thick, black\\%
    ultra thick, blue!60!black, mark=triangle*\\%
    ultra thick, red!70!black, mark=*\\%
    ultra thick, red!70!black, mark=diamond*\\%
    ultra thick, densely dotted, black\\%
    ultra thick, densely dotted, blue!60!black, mark=triangle*, mark options=solid\\%
    ultra thick, densely dotted, red!70!black, mark=*, mark options=solid\\%
}
\usepackage{listings}
\usepackage{wrapfig}
\usepackage{algorithm2e}

\begin{document}

\title{Source-to-Source Automatic Differentiation of OpenMP Parallel Loops}

\author{Jan Hückelheim}
\authornote{Both authors contributed equally to this research.}
\email{jhueckelheim@anl.gov}
\orcid{0000-0003-3479-6361}
\affiliation{%
  \institution{Argonne National Laboratory}
  \streetaddress{9700 South Cass Avenue}
  \city{Lemont}
  \state{Illinois}
  \postcode{60439}
}

\author{Laurent Hascoët}
\authornotemark[1]
\affiliation{%
  \institution{Inria Sophia-Antipolis}
  \streetaddress{2004 Route des Lucioles}
  \city{Biot}
  \country{France}
  \postcode{06902}}
\email{laurent.hascoet@inria.fr}

\begin{abstract}
    This paper presents our work toward correct and efficient automatic differentiation of OpenMP parallel worksharing loops in forward and reverse mode. Automatic differentiation is a method to obtain gradients of numerical programs, which are crucial in optimization, uncertainty quantification, and machine learning. The computational cost to compute gradients is a common bottleneck in practice. For applications that are parallelized for multicore CPUs or GPUs using OpenMP, one also wishes to compute the gradients in parallel. 
    We propose a framework to reason about the correctness of the generated derivative code, from which we justify our OpenMP extension to the differentiation model.
    We implement this model in the automatic differentiation tool Tapenade and present test cases that are differentiated following our extended differentiation procedure. Performance of the generated derivative programs in forward and reverse mode is better than sequential, although our reverse mode often scales worse than the input programs.
%GAIL - What results?
\end{abstract}

\begin{CCSXML}
<ccs2012>
<concept>
<concept_id>10002950.10003714.10003715.10003748</concept_id>
<concept_desc>Mathematics of computing~Automatic differentiation</concept_desc>
<concept_significance>500</concept_significance>
</concept>
<concept>
<concept_id>10011007.10011006.10011041.10011047</concept_id>
<concept_desc>Software and its engineering~Source code generation</concept_desc>
<concept_significance>500</concept_significance>
</concept>
<concept>
<concept_id>10003752.10003753.10003761.10003762</concept_id>
<concept_desc>Theory of computation~Parallel computing models</concept_desc>
<concept_significance>300</concept_significance>
</concept>
<concept>
<concept_id>10003752.10003809.10010170.10010171</concept_id>
<concept_desc>Theory of computation~Shared memory algorithms</concept_desc>
<concept_significance>300</concept_significance>
</concept>
</ccs2012>
\end{CCSXML}

\ccsdesc[500]{Mathematics of computing~Automatic differentiation}
\ccsdesc[500]{Software and its engineering~Source code generation}
\ccsdesc[300]{Theory of computation~Parallel computing models}
\ccsdesc[300]{Theory of computation~Shared memory algorithms}

\keywords{Automatic Differentiation, OpenMP, Shared-Memory Parallel, Multicore}

\maketitle

\section{Introduction}
\label{sec:introduction}

\begin{wrapfigure}{r}{0.4\textwidth}
  \centering
  \vspace{-4em}
    \includegraphics[width=0.9\linewidth]{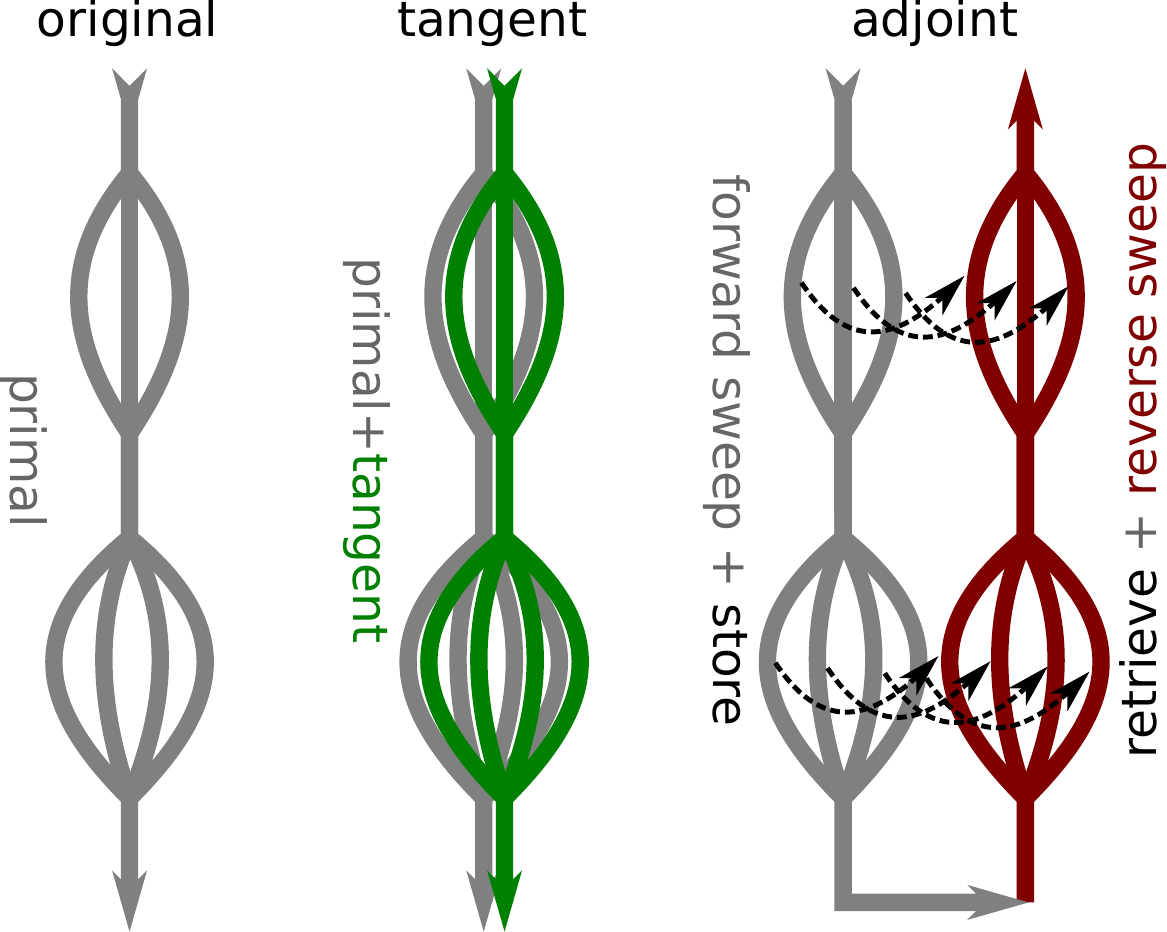}
    \caption{OpenMP worksharing loops use a fork-join parallelism model. Automatic Differentiation can use the same model in forward mode, but the reverse mode poses additional implementation challenges that are discussed in this paper.}
    \vspace{-4em}
    \label{fig:title}
\end{wrapfigure}

Automatic differentiation (AD), also known as algorithmic differentiation or differentiable programming, and the closely related method of backpropagation have in recent years received growing interest for their many uses in optimization, uncertainty quantification, and machine learning. Given a program that implements some mathematical function, AD creates a new program that implements the derivative, gradient, or Jacobian of that function. Many different tools and approaches have been developed to this end; we discuss some of them in Section~\ref{sec:related}.

We focus on \emph{source-transformation AD}, a method in which a given program is transformed into its derivative program before compile time, often resulting in higher efficiency compared with other approaches. Efficiency, in terms of both memory footprint and compute cost of the generated derivative program, is crucial for many AD applications and often requires effective use of parallel computing hardware. While AD for distributed-memory parallelism has been an active research subject for a while  and libraries for adjoint MPI are used in some practical applications, to our knowledge no prior literature describes fully automated source-transformation AD of C or Fortran programs containing OpenMP pragmas or any other shared-memory parallelism, in either tangent or adjoint mode. 

In this work, we present an extension to the Tapenade automatic differentiation tool that supports the differentiation of programs containing OpenMP parallel loops in both forward- and reverse-mode (also known as tangent or adjoint mode), as illustrated in Figure~\ref{fig:title}. This extension is based on a theoretical framework that we propose in order to reason about the correct scoping (e.g., private, shared) of variables in the generated derivative code. For reverse-mode AD, we also present a runtime support library that correctly reverses OpenMP schedules (e.g., static, dynamic) and handles the storage of inputs to nonlinear intermediate operations during the original computation and retrieval during the derivative computation in a thread-safe manner. This extension is now part of the Tapenade master branch and will be available in the next release.

We evaluate our implementation on three test cases: a stencil loop, which follows a computational pattern occurring in structured-mesh solvers, image processing, neural networks, and many other applications; a lattice-Boltzmann method (LBM) solver from the Parboil benchmark suite~\cite{stratton2012parboil}, with applications in computational fluid dynamics; and a Green's function Monte Carlo (GFMC) kernel from the CORAL benchmark suite~\cite{vazhkudai2018design} modeling nuclear physics.

Our paper makes the following contributions:
\begin{itemize}
    \item We present the first theoretical model for the direct differentiation of OpenMP parallel worksharing loops in C/C++ or Fortran. Previous work~\cite{forster2014algorithmic} has considered OpenMP pragmas but only within a simplified language called SPL that lacked for-loops and many other features and did not consider schedules.
    \item We present the first publicly documented general-purpose AD implementation that supports OpenMP. TAF~\cite{giering2003applying} supports some OpenMP, but the differentiation model for OpenMP is closed-source and not publicly documented in the literature. SPLc~\cite{forster2014algorithmic} supports OpenMP but only for SPL programs.
    \item We present performance test cases that compare implementation strategies for shared-memory parallel adjoint derivatives with potential write conflicts. While previous work~\cite{forster2014algorithmic,ssmp,tfmad} has discussed those strategies, there has been a lack of experiments to guide that choice for a user in a general AD setting.
    \item We present the first OpenMP AD runtime library that supports thread-safe stack access and OpenMP schedule reversal for reverse-mode AD.
\end{itemize}

Our work has the following limitations:
\begin{itemize}
    \item We focused exclusively on OpenMP parallel worksharing loops. Some additional pragmas, such as barrier, master, or single, seem like a straightforward extension to our implementation and have been described in theory~\cite{forster2014algorithmic}. Others, such as tasks or target offloading, would require more effort.
%GAIL - in the next item you use the word primal -- in fact you do so elsewhere in the paper but sometimes add a noun suchas variable or code. Can primal stand alon?
    \item We assume, but do not check, that the input program does not contain unspecified behavior. Differentiation may reveal parallelism bugs that were present but did not manifest in the primal because of implementation-defined behavior. Checking for input code correctness is impossible in general and certainly beyond the scope of our work.
    \item The current Tapenade implementation supports OpenMP only in Fortran. We expect C support to be a straightforward extension to the OpenMP parser, but leave this for future work. C++ (even without OpenMP) is not currently supported by Tapenade at all and would require substantial development and research effort.
    \item Concurrent read access in the primal leads to concurrent increments, which our implementation turns into atomic updates or reductions. While some parallel speedup still results, the adjoint codes in our experiments scale worse than the input codes, which is expected in the general case, and can only be improved upon in special cases. This has also been observed in other work for SPL~\cite{forster2014algorithmic} and Fortran~\cite{ssmp,tfmad}. Future work should address this by increasing the number of cases in which atomics/reductions can be avoided, and/or by implementing faster methods for parallel reductions.
\end{itemize}

Our paper is structured as follows.
In Section~\ref{sec:background}
we briefly summarize OpenMP and AD concepts that are relevant to this work.
In Section~\ref{sec:model}
we explain our theoretical framework
and its implementation,
and in Section~\ref{sec:implementation}
we discuss the implementation  of the runtime support library.
In Section~\ref{sec:testcases}
we present  our test cases and performance measurements
In Section~\ref{sec:related}
we discuss related work, and
in Section~\ref{sec:conclusion}
we review our results.

\section{Background}
\label{sec:background}
In this section we summarize OpenMP programming and AD concepts that are relevant to this work. Even though we expect most readers to be familiar with at least basic OpenMP programming, we highlight some subtleties that affect the differentiation of such programs.

\subsection{OpenMP}
\label{sec:openmp}
OpenMP is an industry standard (version 5.1 is the most recent at the time of writing) that allows portable parallelization of C, C++, and Fortran programs, mostly for shared-memory architectures. Recent extensions include tasks, SIMD vectorization, and offloading to GPUs and other accelerators. In this work, we restrict ourselves to one of the most widely used features, which has been part of the OpenMP standard from the first version, namely the parallelization of worksharing loops~\cite{dagum1998openmp}. A \texttt{for}-loop (C/C++) or \texttt{do}-loop (Fortran) that satisfies certain structural requirements, in OpenMP language called \emph{canonical loop form}, can be parallelized by placing an OpenMP \emph{worksharing loop construct} in front of the loop header. An example OpenMP program is shown in Figure~\ref{fig:sampleomp}.

This causes the loop iteration space to be split into \emph{chunks}, which are then distributed among the available threads. The size of the chunks, as well as the assignment of chunks to threads, is controlled by the OpenMP \emph{schedule}. The user may explicitly choose a schedule using environment variables or runtime library calls or by adding a \texttt{schedule} clause to the worksharing loop construct. The OpenMP standard defines a number of schedules, including a \texttt{static} schedule, which assigns roughly equal parts of the iteration space to the threads a priori in a determinstic fashion, as well as \texttt{dynamic} and \texttt{guided} schedules, which use different strategies to improve load balancing by dynamically adjusting the assignment of chunks to threads at runtime.

In addition, OpenMP provides the \emph{scoping} clauses \texttt{shared}, \texttt{private}, \texttt{firstprivate}, \texttt{lastprivate}, and \texttt{reduction}. Scoping clauses contain a list of variables to which the clause is applied. OpenMP also has a default scope for variables that are not listed in any clause, which is usually \texttt{shared}, with some exceptions. All clauses except \texttt{shared} will result in a process called \emph{privatization}. For each variable that is privatized, a local version of that variable is declared on each thread, which is accessible instead of the original variable during the parallel region. Note that this can cause programs containing OpenMP pragmas to behave differently depending on whether or not the compiler supports OpenMP (or has OpenMP support activated, for example by using a command line option such as \texttt{-fopenmp}).
% Consider as an example the following program.
%\begin{lstlisting}
%a = 1
%omp parallel do private(a)
%do i=1,10
%  a = 0
%end do
%print *, a
%\end{lstlisting}
If the program in Figure~\ref{fig:sampleomp} is interpreted without OpenMP support, the pragma is treated as a comment. The reference \texttt{a} refers to the same variable inside and outside the loop, and hence the value of \texttt{a} is first set to 1, then 100 times overwritten with 0, and the program will finally print "0", followed by a printout of the array \texttt{arr}, which is "1" followed by many zeros. In contrast, an OpenMP-capable compiler must privatize \texttt{a} inside the parallel region, which means that the reference \texttt{a} inside the loop refers to a separate variable on each thread, all of which are separate from the original variable that is accessible outside the parallel loop. The program must therefore print "1", followed by a printout of \texttt{arr} that in this case contains "1" in each index that was the first to be modified on any thread, and "0" otherwise. Note that the original and all private entities of \texttt{a} exist in memory at the same time, which can drastically increase the memory footprint of the program if \texttt{a} is a large object or array, particularly if the number of threads is high. Since the dataflow of a program may change depending on whether or not OpenMP support is active in the compiler, our AD implementation also has a command line switch to that effect.

\begin{figure}
    \centering
    \begin{lstlisting}
program test
real arr(100)
a=1
!$omp parallel do firstprivate(a) shared(arr) schedule(dynamic)
do i=1,100
  arr(i) = a
  a = 0
end do
write(*,*) a, arr
end program test
\end{lstlisting}
    \caption{Example OpenMP program. Even though this program is legal, it has a data flow that causes non-determinstic results in \texttt{arr} that also depend on the number of threads, and a different result in \texttt{a} depending on whether the program is compiled with or without OpenMP support. Our method enables the correct differentiation of such general programs.}
    \label{fig:sampleomp}
\end{figure}

The privatization scopes differ in the way that the variables are influenced by, or influence, their corresponding original variable. For example, the privatized instances of a variable declared \texttt{firstprivate} will be initialized with the value of the corresponding original variable, while variables with \texttt{reduction} scope will be initialized to a neutral element, and the thread-local instances are combined with the original variable using the reduction operator. Apart from the initialization and combination, the reduction clause behaves just like any other privatization clause, and a program therefore can use or modify a reduction variable in arbitrary ways.

The scoping clauses must be taken into account during differentiation, since they affect the dataflow and the semantics of the program, as we explain in more detail in Section~\ref{sec:model} and illustrate in Figure~\ref{fig:privateflow}. We note that a parallel loop may have dataflow across iterations. In particular, the lifetime of private variables on each thread spans an entire parallel region, not just a loop iteration. A thread may, for example, allocate a private data structure in its first iteration, read, update, or reuse it across multiple loop iterations, and then deallocate it in its last iteration.

Because private variables stay alive across iterations within one parallel loop, the OpenMP schedule may also affect the dataflow and program semantics. This is also shown in Figure~\ref{fig:sampleomp}. Because each thread copies the value of the firstprivate variable \texttt{a} into one particular index in \texttt{arr} before overwriting its own private instance of \texttt{a}, the shared array \texttt{arr} will be zero except at the indices that correspond to the first iteration that each thread was assigned, resulting in a schedule-dependent dataflow.

Sometimes one wishes to keep private variables alive across multiple parallel regions. OpenMP supports this feature through the \texttt{threadprivate} pragma, which can be placed before the declaration of certain variables (the detailed restrictions are given in the standard). We use this feature for our AD implementation to store intermediate results during the original computation and retrieve them in the corresponding derivative computation. This requires that both computations occur on the same thread, which must be ensured by our AD tool regardless of the OpenMP schedule.

\subsection{Source-Transformation AD}
\label{sec:stsAD}

Here we aim to present the subset of Automatic Differentiation theory and the notations used in the rest of the paper. A complete introduction to AD can be found in e.g. \cite{Griewank2008EDP} or \cite{Giering1996RfA}.
Automatic differentiation identifies a sequence of instructions performed by a given (\emph{primal}) program
\begin{equation}
{\tt P:}\>\{I_1; I_2; \dots I_p;\}
\label{eqn:prginstrs}
\end{equation}
with a composition of mathematical functions
\begin{equation*}
F = f_p \circ f_{p-1} \circ \dots \circ f_1\enspace.
\end{equation*}
Strictly speaking, each program variable possibly represents several mathematical variables, one for each time the program variable is overwritten. Each instruction $I_k$ is identified to a function $f_k$ that operates on this larger set $X$ of mathematical variables. Each $I_k$ thus applies $f_k$ to the output of $f_{k-1}$. If the program contains control flow, the instruction sequence in \eqref{eqn:prginstrs} is that of the executed code path. By the chain rule of calculus, the derivative of $F$ is
\begin{equation*}
F'(X) = f'_p(X_{p-1})\> \times\> f'_{p-1}(X_{p-2})
\> \times\> \dots\> \times\> f'_1(X_0) \enspace,
\end{equation*}
where $X_0=X$ and $X_k=f_k(X_{k-1})$ for $k=1$ to $p$.
Practically, codes produced by AD will compute the product of $F'$ with an appropriate seed vector, either on the left or on the right. This approach is more efficient for applications that only require Jacobian-vector products. The full $F'(X)$ can be obtained through repeated evaluation of the derivative code with all vectors of the Cartesian basis as seeds.

With a seed column vector $\dot{X}$ multiplied on the right, one obtains the \emph{tangent mode} of AD,
\begin{equation*}
\dot{Y} = F'(X)\times\dot{X} = f'_p(X_{p-1})\> \times\> f'_{p-1}(X_{p-2})
\> \times\> \dots\> \times\> f'_1(X_0) \>\times\> \dot{X}\enspace,
\end{equation*}
which computes the directional derivative of $F$ along direction $\dot{X}$. The efficient computation order is right to left. This results in a \emph{tangent program} that evaluates the $f'_k$ from $k=1$ to $p$, that is, in the same order as the primal program. Therefore the tangent program built by source-transformation AD retains the general control structure of the primal program, with each primal instruction accompanied (actually preceded) by its derivative instruction.

With a seed row vector $\overline{Y}$ multiplied on the left, one obtains the \emph{adjoint mode} of AD,
\begin{equation*}
\overline{X} = \overline{Y}\times F'(X) = \overline{Y}\> \times\> f'_p(X_{p-1})\> \times\> f'_{p-1}(X_{p-2})
\> \times\> \dots\> \times\> f'_1(X_0)\enspace,
\end{equation*}
which computes the gradient with respect to $X$ of the scalar function $\overline{Y}\times F(X)$. The efficient computation order is left to right. This results in an  \emph{adjoint program} that evaluates the $f'_k$ from $k=p$ down to $1$, that is, in the inverse order of the primal program. This is certainly more complicated than the tangent program but has an enormous benefit: in the frequent case of a program with many inputs and few or just one output (think of a simulation leading to a scalar-valued cost function) this will return the complete gradient of the function with respect to all of its inputs in just one run. This benefit largely compensates for the resulting extra complexity, with a source-transformation adjoint code made of one \emph{forward sweep} followed by a \emph{backward sweep}. Since the derivatives are computed in the reverse order, the code that evaluates them (the \emph{backward sweep}) must reverse all control structures of the primal program. This is called control-flow reversal. Moreover, since each $f'_k$ is evaluated at point $X_{k-1}$, all the $X_k$ must be provided to the backward sweep in the inverse of their primal computation order. This is called dataflow reversal. For both reasons, our adjoint AD model chooses to precede the backward sweep with a \emph{forward sweep}, which is basically a copy of the primal code equipped with storage of the intermediate values and control decisions that will be used in the backward sweep. The ``mirror-like'' correspondence between the forward and backward sweeps indicates that a last-in first-out stack datastructure can naturally serve as a storage mechanism for this application. It should be noted that instead of storing all intermediate values, it is often possible to mix storage and recomputation from nearby stored values, allowing tradeoffs between computational cost and memory footprint that are beyond the scope of this work. An approach that completely eliminates storage in favor of recomputation would remove the need for the thread-safe stack implementation discussed in this work, but at a computational cost that is infeasible for many real-world applications. In the remainder of this work we will therefore assume a differentiation procedure that stores intermediate values, as implemented in Tapenade.

In Section~\ref{sec:model} we will look in more detail at the derivative instructions in both tangent and adjoint programs. Derivative code for function or procedure calls is mainly a technical issue, related to the particular AD model chosen by an AD tool, and is irrelevant here. We will concentrate on assignments, putting aside other kinds of instructions (I/O, \dots), whose derivative are nonexistent or trivial. Consider any particular primal scalar assignment $I_k$
\begin{center}
    {\tt z = x $\hbox{\it op}_k$ y}
\end{center}
with $\hbox{\it op}_k$ any differentiable (here binary) operation. Let us assume for clarity, without loss of generality, that the memory location of {\tt z} does not overlap with any variable in the right-hand side. If not, just introduce a temporary variable and split the assignment.
The assignment's tangent derivative is
\begin{center}
{\tt $\dot{\tt z}$ = ${{\partial \hbox{\it op}_k}\over {\partial {\tt x}}}({\tt x},{\tt y})$ * $\dot{\tt x}$ +  ${{\partial \hbox{\it op}_k}\over {\partial {\tt y}}}({\tt x},{\tt y})$ * $\dot{\tt y}$}.
\end{center}
The adjoint derivative implements a less intuitive vector$\times$matrix product, resulting here in the sequence of assignments
\begin{center}
\begin{minipage}[t]{.3\linewidth}
    {\tt $\overline{\tt x}$ = $\overline{\tt x}$ + $\overline{\tt z}$ * ${{\partial \hbox{\it op}_k}\over {\partial {\tt x}}}({\tt x},{\tt y})$}\\
    {\tt $\overline{\tt y}$ = $\overline{\tt y}$ + $\overline{\tt z}$ * ${{\partial \hbox{\it op}_k}\over {\partial {\tt y}}}({\tt x},{\tt y})$}\\
    {\tt $\overline{\tt z}$ = 0}.
    \end{minipage}
\end{center}
We observe that variable {\tt x} (resp. {\tt y}) being read results in variable $\overline{\tt x}$ (resp. $\overline{\tt y}$) being incremented. Notice also that $\overline{\tt z}$ is reset to zero. While this follows naturally from the vector$\times$matrix multiplication, one can support intuition considering that any $\overline{\tt v}$ is the {\em influence} of {\tt v} on the final result, at the corresponding location in the primal code. For instance the resulting $\overline{\tt z}$ is the influence of the {\tt z} immediately before $I_k$, on the final result. It is obviously zero as {\tt z} is about to be overwritten. We will also need to single out the special case of a primal assignment that just {\em increments} variable {\tt v}. One can show that the resulting adjoint instruction(s) only read $\overline{\tt v}$. Figure~\ref{fig:elementaryadj} summarizes the access patterns occurring in the backward sweep of the adjoint code, for each relevant access pattern of a variable {\tt v} in the primal code i.e. pure write, pure read, increment, coming into scope (shown as setting to undefined ${\tt v}=\_$), and falling out of scope (shown as ${\tt v}\!\!\!/$). It also shows the {\tt POP} calls that restore a previous value of {\tt v}.
\begin{figure}
\begin{center}
{\small
\begin{tabular}{|l|l|l|l|l|}
     \hline
     ${\tt v}=...; $ &
     $=...{\tt v}; $ &
     ${\tt v}+\!\!=...; $ &
     ${\tt v}=\_$ &
      ${\tt v}\!\!\!/$\\
     \hline
     $\hbox{\tt POP}({\tt v});\>\> =...\overline{\tt v};\>\>\overline{\tt v} = 0;$ &
     $\overline{\tt v}+\!\!=...;\>\>=...{\tt v}; $ &
     $\hbox{\tt POP}({\tt v});\>\> =...\overline{\tt v};$&
     $\overline{\tt v}\!\!\!/ \>\>\> {\tt v}\!\!\!/$&
     $\hbox{\tt POP}({\tt v});\>\> \overline{\tt v} = 0;$\\
     \hline
\end{tabular}
}
\end{center}
\caption{Access patterns of the adjoint and primal variables in the adjoint code (bottom) corresponding to relevant access patterns of the primal variable in the primal code (top). Access patterns ${\tt v}=\_$ and ${\tt v}\!\!\!/$ are only pseudo-code standing for {\tt v} coming into and falling out of scope, but the latter causes actual adjoint code to be generated.}
\label{fig:elementaryadj}
\end{figure}

We remark that both tangent AD and, even more, adjoint AD make use of several auxiliary variables, for instance to implement reversal of the control flow or to eliminate common subexpressions introduced by differentiation. When inside an OpenMP parallel region, the AD tool must (and our implementation will) make sure these auxiliary variables do not conflict across threads, by assigning them the {\tt private} scoping.

\section{Differentiation Model for OpenMP Code}
\label{sec:model}

Given a primal code implemented with OpenMP, we describe a way for an AD tool to deduce a correct and efficient OpenMP-parallel differentiated code. The answer, immediate for tangent differentiation, is less obvious but more interesting for adjoint differentiation, considering the dataflow reversal that adjoint AD requires. We note that there are of course (infinitely) many ways to implement any given function, and that our differentiation strategy is thus only one possible option. We will point out interesting alternatives that we are aware of throughout the text when appropriate.

As seen in Section~\ref{sec:openmp}, the OpenMP model for worksharing loops works mostly by designating a parallel region and by specifying, for designated loops in this region, a scoping for each variable, that is, either \texttt{shared} between threads or \texttt{private} to each thread, possibly with variants (e.g., \texttt{firstprivate}, \texttt{lastprivate}, or \texttt{reduction}).
Our problem thus naturally splits into two questions. First, given a primal code that we will assume correct (no race conditions or unspecified behavior), what regions of the differentiated code can be declared parallel? Second, in these differentiated parallel regions, what is the best scoping that can be attached to each variable, knowing that a variable in the differentiated code can be either the replica or the derivative of some primal variable?

Notice that for any array reference in the primal code, which uses some indices, both the corresponding primal copy and derivative reference in the differentiated code are array references with the same indices. This is one of the keys to guaranteeing the equivalent shapes of the data-dependency graphs that we will use in the next section.

\subsection{Parallel Regions of Differentiated Code}
\label{sec:modelRegions}

The structure of a differentiated code follows that of the primal code, as described in Section~\ref{sec:stsAD}. For tangent AD, a clear correspondence exists from any region or control-flow structure of the primal code to its tangent counterpart, which interleaves primal and derivative computations using both primal and differentiated variables. For adjoint AD, any primal region has two corresponding regions: one in the forward sweep and one in the backward sweep. The forward corresponding region essentially repeats instructions of the primal code and saves intermediate values and controls on a special stack. The backward corresponding region computes derivatives, thus writing in differentiated variables and reading from both differentiated and primal variables, whereas primal variables are overwritten only when restoring their previous value, popped from the stack. We will show that the parallel regions of the primal code can be propagated to the differentiated code, that is, to the corresponding region of the tangent code as well as to both corresponding regions of the adjoint code. We note that in some cases it would be possible to fuse a parallel region of the forward sweep and a parallel region of the reverse sweep into just one parallel region, which may reduce run time and peak memory consumption. However, this depends on the position of these parallel regions in the code, and requires a detailed analysis of loop dependencies, which may very well exist even in OpenMP-parallel code as discussed in Section~\ref{sec:background}.

\subsubsection{Tangent-Linear}

The case is clear for the tangent code, whose skeleton is basically a copy of the primal code with an added differentiated instruction inserted when needed before each primal assignment. The data-dependency graph of the tangent code is the union of three subgraphs:
\begin{itemize}
    \item Dependencies between primal variables, which form a copy of the primal code dependency graph, sometimes even smaller thanks to possible simplification such as slicing, where program parts that do not affect derivatives are removed
    \item Dependencies between differentiated variables, which form a graph isomorphic to the primal dependency graph, sometimes even smaller when some variables have no differentiated counterpart
    \item Dependencies from read primal variables to written differentiated variables, occurring due to any nonlinear operation of the primal code, and that follow the same pattern as the primal code dependencies from right-hand side reads to left-hand side writes.
\end{itemize}
Therefore, the complete tangent data-dependency graph inherits the shape and properties of the primal and, in particular, dependency distances over the iteration space of loops. Tangent differentiated loops thus inherit the parallel status of primal loops. Likewise, tangent differentiated parallel sections remain independent so they can also be declared as parallel sections.

\subsubsection{Adjoint}

Regarding the adjoint code, the forward sweep is also basically a copy of the primal code, with some instructions possibly simplified out and, more important, with added ``{\tt PUSH}'' instructions that save intermediate values. The data-dependency graph of the forward sweep is again a copy or a subgraph of the primal graph, plus dependencies relating writes of intermediate values and stack {\tt PUSH} operations. Provided we adapt the stack mechanism (see Section~\ref{sec:stack}) to prevent any cross-thread dependency on the stack itself, the forward sweep of a primal parallel region, loop, or section inherits the status of the primal region.

Analysis of the backward sweep requires that we take into account the dataflow reversal inherent to adjoint differentiation. Before that, we need to introduce a refinement to our data-dependency graphs to take into account a remarkable property of {\em increment} operations: just as two successive {\em reads} of a given variable {\tt v} introduce no ordering constraint and therefore there is no data dependency between two {\em reads}, we observe that the same property holds between two successive {\em increments} of {\tt v} (at least in exact arithmetic), namely, operations of the form {\tt v {\tt +=} expression}. To take advantage of this feature, we introduce a new type of node in data-dependency graphs: in addition to the classical read and write nodes, we define increment nodes collapsing into a single node the read and write nodes that the increment consists of. In doing so, we assume that each increment is actually {\em atomic}, which is not guaranteed in an execution model such as OpenMP. We will show in Section~\ref{sec:modelVars} how we ensure atomicity of increment operations on adjoint differentiated variables. The classical building rules for data dependencies are easily extended by stating that no dependency exists between two successive increments, whereas dependencies do  exist between successive accesses increment-to-write, write-to-increment, increment-to-read, and read-to-increment.

Going back to adjoint differentiation, we first observe that the backward sweep accesses both primal variables and derivative variables. Primal variables may be read only by the derivative of code that read them in the primal. They may be overwritten only through ``{\tt POP}'' operations from the stack. Our AD model places each ``{\tt POP}'' at the symmetric location of its ``{\tt PUSH,}'' which is itself immediately before the overwrite of the primal variable (\emph{save-on-kill}). Therefore the subgraph of the adjoint data-dependency graph that deals with primal variables is a reversed copy of the primal data-dependency graph. Regarding the subgraph that deals with derivative variables, we observe (see Figure~\ref{fig:elementaryadj}) a one-to-one correspondence from the operations of the primal code on primal variables to the operations of the adjoint code on the adjoint variables, only in exact reverse order. Consider any primal code assignment and any variable reference {\tt v} appearing in the assignment:
\begin{itemize}
    \item The assignment may {\em read} {\tt v} one or more times and never overwrite it. It follows from the adjoint AD model that its adjoint code will only {\em increment} the derivative $\overline{\tt v}$ one or more times.
    \item The assignment may {\em increment} {\tt v} and not access {\tt v} otherwise. It follows from the adjoint AD model that its adjoint code will only {\em read} the derivative $\overline{\tt v}$
    \item In all other cases, the assignment may {\em read} {\tt v} zero or more times, then finally overwrite {\tt v}. It follows from the adjoint AD model that its adjoint code will {\em read} $\overline{\tt v}$ and finally overwrite it.
\end{itemize}
Since {\em read} and {\em increment} operations play symmetrically the same role in data dependency, the data-dependency graph of the backward sweep is (a subgraph of) a reversed copy of the primal graph. This shows that the parallel status of a code fragment is transferred to its corresponding backward sweep.

In summary, the parallel regions, loops, and sections of the primal code can transfer their parallel status to their counterpart regions, loops, and sections of the tangent differentiated code and likewise to both the forward sweep and the backward sweep of the adjoint differentiated code. However, this capability will bear fruit only when we have decided which scoping we can attach to each variable, primal copy and differentiated, that appears in the differentiated parallel regions. This topic is studied in the next section.

\subsection{Scoping of Variables of the Differentiated Code}
\label{sec:modelVars}

For each variable {\tt v} of the primal code that appears in a parallel region, and knowing its scoping ({\tt shared}, {\tt private}, \dots) in the primal code, we want to find the scoping we may attach to the primal copy and to differentiated counterparts of {\tt v} in the differentiated parallel regions, in order  to allow for correct and efficient parallel execution.

In the following analysis, if the variable of interest is an array, our notation {\tt v} symbolizes one particular array cell, and we consider only the accesses that (may) affect this cell. Since the OpenMP clauses can be attached only to the variable name, we must make sure in the end that our proposed scoping for each cell of the array is the same, and this will be the scoping attached to the variable name itself in the OpenMP clauses. This applies to the primal copy of {\tt v} in the differentiated code as well as to its derivative $\dot{\tt v}$ or $\overline{\tt v}$

\subsubsection{Tangent-Linear}

Here also tangent differentiation is the easy case. Jumping to the conclusion, the primal copy of any {\tt v} can receive exactly the scoping {\tt v} has in the primal code. The same holds for its tangent derivative $\dot{\tt v}$. As an exception, {\tt reduction} scoping other than sum cannot be extended directly to $\dot{\tt v}$. We leave this to future work.

We will show the validity of these transformations on a few tangent differentiation cases only, as a training ground for the justification diagrams that we will use later for adjoint differentiation. We will use the PGAS transformation \cite{refPGAS}, which rewrites a piece of parallel code, renaming variables that bear the same name but are distributed (privatized) over different threads into variables with different names. This transformation results in an equivalent code that operates on shared memory only. One can then demonstrate properties of this code or, in our case, apply some semantically valid transformations before trying to apply reciprocal memory conversion to reintroduce thread-local memory. Therefore our justification diagrams are built of four sections. The first is the primal code of some OpemMP parallel region, the second is its PGAS translation, the third is the derivative code of the PGAS translation, and the fourth is an OpenMP code whose PGAS translation is the contents of the third section.
In each section we consider only the accesses to a variable {\tt v} of interest. The parallel region must be thought of as embedded in a larger code on which we make no assumption, and for sake of clarity we will not show it in the diagrams. To capture all possible data-dependencies, just assume (for the top two sections) that the primal parallel region is preceded by access patterns ${\tt v}=...;\>\> =...{\tt v};$ and is followed by $=...{\tt v};\>\>{\tt v}=...;$. Consequently the code in the last two sections is preceded by the corresponding tangent access pattern $\dot{\tt v}=...;\>\> {\tt v}=...;\>\> =...\dot{\tt v};\>\>=...{\tt v};$ and likewise is followed by $=...\dot{\tt v};\>\>=...{\tt v};\>\>\dot{\tt v}=...;\>\> {\tt v}=...;$ independently of the OpenMP pragmas applied to {\tt v} in the parallel region.

Consider, for instance, a parallel region that declares one variable {\tt v} as {\tt private}. The four steps of the validation process are shown in the diagram of Figure~\ref{fig:privatetgt}. In the primal code of the top section, we make no assumption about the accesses to {\tt v}, resulting in any sequence of read or write accesses to {\tt v}, denoted as $[ {\tt v}=... \>|\>  =...{\tt v} ]*$.
The second section shows the equivalent PGAS-style code where each {\tt v} inside the parallel region is replaced with a new {\tt v}$_T$ for each particular thread. The original {\tt v} still exists and retains its value throughout the parallel region. Each {\tt v}$_T$ is initially undefined (denoted by ${\tt v}_T=\_$) since {\tt v} is not {\tt firstprivate}. Likewise, {\tt v}$_T$ falls out of scope (denoted by ${\tt v}\!\!\!/_T$) at the closing of the thread. We then apply tangent differentiation, yielding the third section. We observe that the third section is exactly the PGAS-style code that we would obtain from the particular standard OpenMP code shown in the bottom section.

\begin{figure}
\begin{center}
{\small
\begin{tabular}{|l|}
     \hline
     $\hbox{\tt \$OMP PARALLEL PRIVATE(v)}\>\> \{ $\\
     $\>\>\>\>[ {\tt v}=...; \>|\> =...{\tt v}; ]*\>\> \} $\\
     \hline
     $\forall\>\hbox{\em thread}\>T\>\> \{ $\\
     $\>\>\>\>{\tt v}_T=\_;\>\> [ {\tt v}_T=...; \>|\> =...{\tt v}_T; ]*\>\>\>\>{\tt v}\!\!\!/_T\> \} $\\
     \hline
     $\forall\>\hbox{\em thread}\>T\>\> \{ $\\
     $\>\>\>\>\dot{\tt v}_T=\_;\>\> {\tt v}_T=\_;\>\> [ \dot{\tt v}_T=...;\> {\tt v}_T=...; \>|\> =...\dot{\tt v}_T;\> =...{\tt v}_T; ]*\>\>\>\>{\tt v}\!\!\!\dot{/}_T\>\>{\tt v}\!\!\!/_T\> \} $\\
     \hline
     $\hbox{\tt \$OMP PARALLEL PRIVATE($\dot{\tt v}$,v)}\>\> \{ $\\
     $\>\>\>\>[ \dot{\tt v}=...;\>\> {\tt v}=...; \>|\> =...\dot{\tt v};\>\> =...{\tt v}; ]*\>\> \} $\\
     \hline
\end{tabular}
}
\end{center}
\caption{Validation diagram for tangent AD of the \texttt{private} scoping. Notice the pseudo-code shown in the PGAS sections for ${\tt v}_T$ coming into and falling out of scope. Both {\tt v} and $\dot{\tt v}$ can inherit {\tt v}'s private scoping.}
\label{fig:privatetgt}
\end{figure}

Figure~\ref{fig:sharedtgt} validates tangent differentiation of the \texttt{shared} scoping. Reference {\tt v} is not expanded any more into several {\tt v}$_T$. Assuming that the primal program is correct (i.e., deterministic), only three cases arise. First, the memory cell of {\tt v} may be accessed by only one thread (left diagram), in which case there is no restriction on the kind of access (read, write , \dots). Second, the memory cell of {\tt v} is accessed by possibly many threads but only for {\em reading} (middle diagram). Third,  {\tt v} is accessed by possibly many threads but only for {\em atomic} increment operations (denoted as \framebox {\makebox[\totalheight]{\tt +=}}). The tangent-linear code then only increments $\dot{\tt v}$ (right diagram), and these increments must be declared {\tt atomic}, too. In all three cases, diagrams show that a {\tt shared} $\dot{\tt v}$ implements the desired behavior.

\begin{figure}
\begin{center}
{\small
\begin{tabular}{|l|}
     \hline
     $\hbox{\tt \$OMP PARALLEL SHARED(v)}\>\> \{ $\\
     $\>\>\>\>[{\tt v}=...; \>|\> =...{\tt v}; ]*\> \} $\\
     \hline
     $\hbox{\em On only one thread}\>T\>\> \{ $\\
     $\>\>\>\> [{\tt v}=...; \>|\> =...{\tt v}; ]*\> \} $\\
     \hline
     $\hbox{\em On only one thread}\>T\>\> \{ $\\
     $\>\>\>\>[ \dot{\tt v}=...;\> {\tt v}=...; \>|\> =...\dot{\tt v};\> =...{\tt v}; ]*\> \} $\\
     \hline
     $\hbox{\tt \$OMP PARALLEL SHARED($\dot{\tt v}$,v)}\>\> \{ $\\
     $\>\>\>\>[ \dot{\tt v}=...;\>\> {\tt v}=...; \>|\> =...\dot{\tt v};\>\> =...{\tt v}; ]*\> \} $\\
     \hline
\end{tabular}
\hfill
\begin{tabular}{|l|}
     \hline
     $\hbox{\tt \$OMP PARALLEL SHARED(v)}\>\> \{ $\\
     $\>\>\>\>[\> =...{\tt v}; ]*\> \} $\\
     \hline
     $\forall\>\hbox{\em thread}\>T\>\> \{ $\\
     $\>\>\>\>[\> =...{\tt v}; ]*\> \} $\\
     \hline
     $\forall\>\hbox{\em thread}\>T\>\> \{ $\\
     $\>\>\>\>[\> =...\dot{\tt v};\> =...{\tt v}; ]*\> \} $\\
     \hline
     $\hbox{\tt \$OMP PARALLEL SHARED($\dot{\tt v}$,v)}\>\> \{ $\\
     $\>\>\>\>[\> =...\dot{\tt v};\>\> =...{\tt v}; ]*\> \} $\\
     \hline
\end{tabular}
\hfill
\begin{tabular}{|l|}
     \hline
     $\hbox{\tt \$OMP PARALLEL SHARED(v)}\>\> \{ $\\
     $\>\>\>\>[\> {\tt v}\hbox{\framebox {\makebox[\totalheight]{\tt +=}}}...; ]*\> \} $\\
     \hline
     $\forall\>\hbox{\em thread}\>T\>\> \{ $\\
     $\>\>\>\>[\> {\tt v}\hbox{\framebox {\makebox[\totalheight]{\tt +=}}}...; ]*\> \} $\\
     \hline
     $\forall\>\hbox{\em thread}\>T\>\> \{ $\\
     $\>\>\>\>[\> \dot{\tt v}\hbox{\framebox {\makebox[\totalheight]{\tt +=}}}...;\>\> {\tt v}\hbox{\framebox {\makebox[\totalheight]{\tt +=}}}...; ]*\> \} $\\
     \hline
     $\hbox{\tt \$OMP PARALLEL SHARED($\dot{\tt v}$,v)}\>\> \{ $\\
     $\>\>\>\>[\> \dot{\tt v}\hbox{\framebox {\makebox[\totalheight]{\tt +=}}}...;\>\> {\tt v}\hbox{\framebox {\makebox[\totalheight]{\tt +=}}}...; ]*\> \} $\\
     \hline
\end{tabular}
}
\end{center}
\caption{Validation diagrams for tangent AD of the \texttt{shared} scoping.
For a given memory reference {\tt v}, only the three legal cases need be considered: any sort of access to {\tt v} but by only one thread (left), all threads may access but only for reading (middle), all threads may access but only for atomic increments (right). In all three cases, {\tt v} and $\dot{\tt v}$ can receive {\tt v}'s scoping.
}
\label{fig:sharedtgt}
\end{figure}

\subsubsection{Adjoint}

\begin{figure}
    \centering
    \includegraphics[width=\linewidth]{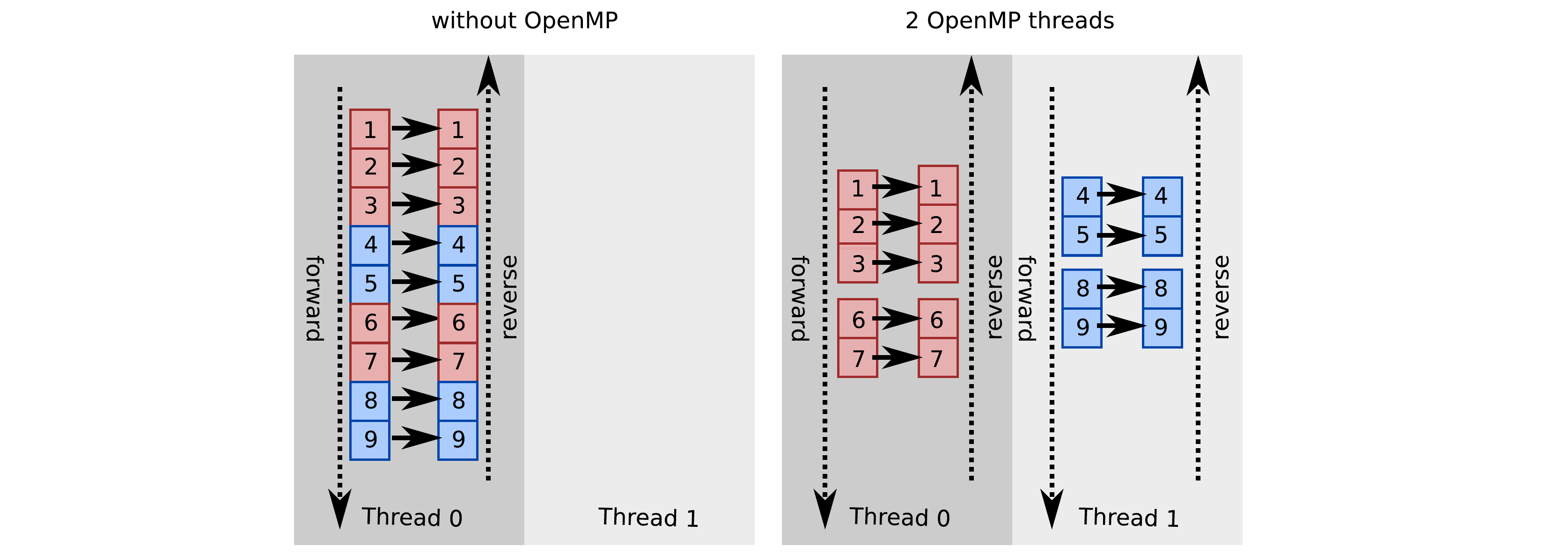}
    \caption{Data dependencies for intermediate results. Each thread has its own stack, and it is important that the same thread will perform the corresponding primal and reverse iterations and perform the reverse operations in the reverse order, to ensure that the correct data is popped from the stack.}
    \label{fig:stack}
\end{figure}

Let us now move to adjoint differentiation. In the forward sweep, just as for tangent differentiation, all variables may retain their scoping from the primal code. A delicate part concerns the stack of intermediate values that is used internally by the ``{\tt PUSH}'' and ``{\tt POP}''. Even if the primal code is indeed parallel and parallel loop iterations or code sections can be executed in any order, the additional need to preserve the {\tt PUSH/POP} correspondence imposes constraints, illustrated by Figure~\ref{fig:stack}.
First, the stack itself must become private to each thread (actually {\tt threadprivate}, see Section~\ref{sec:stack}) because concurrent `{\tt PUSH}'' and ``{\tt POP}'' on the same stack will make the code nondeterministic. Second, parallel thread scheduling in the backward sweep must mirror exactly that of the forward sweep, to make sure that the thread that {\tt PUSH}es a value in a forward parallel loop is the same that will {\tt POP} it during the backward parallel loop. This implies that the forward scheduling be recorded and reused (in reverse) as the backward scheduling (see Section~\ref{sec:schedulereversal}). Only then can we guarantee that values popped are always correct whatever the parallel thread execution order.

We focus now on the backward sweep of adjoint differentiation. In the justification diagrams that follow, the bottom two sections show the {\em backward} sweep of the adjoint code, i.e. in essence the computation of the derivatives plus the {\tt POP} operations that restore primal values. The {\tt PUSH}'es do not appear as they belong to the forward sweep. The adjoint of each instruction is built by applying mechanically the transformation given by Figure~\ref{fig:elementaryadj}, only in reverse order e.g. the beginning of the third section is the adjoint of the end of the second section. In the second section, we explicitly show pseudo-code for a variable coming into scope (${\tt v}=\_$) or falling out of scope (${\tt v}\!\!\!/$) to justify the corresponding adjoint code. 

Since we are now dealing with adjoint differentiation, the code in the last two sections of each diagram must be thought of as preceded by $$\hbox{\tt POP}({\tt v});\>\>  =...\overline{\tt v};\>\> \overline{\tt v}=0;\>\> \overline{\tt v}{\tt +\!\!= }...;\>\>  =...{\tt v}; $$ which is the adjoint of the code pattern ($=...{\tt v};\>\>{\tt v}=...;$) following the primal parallel region, and as followed by $$\overline{\tt v}{\tt +\!\!= }...;\>\>  =...{\tt v};\>\> \hbox{\tt POP}({\tt v});\>\>  =...\overline{\tt v};\>\> \overline{\tt v}=0; $$ which is the adjoint of the pattern (${\tt v}=...;\>\> =...{\tt v};$) preceding the primal parallel region. These adjoint patterns result from mechanical application of the transformation given by Figure~\ref{fig:elementaryadj}, and are the same whatever the scoping studied. We mention these pre- and post-patterns to illustrate the data-dependency context of the adjoint parallel region, which is quite general as it exposes reads and writes of $\overline{\tt v}$, and shows the location of the nearest stack operations.

\begin{figure}
    \centering
    \includegraphics[width=\linewidth]{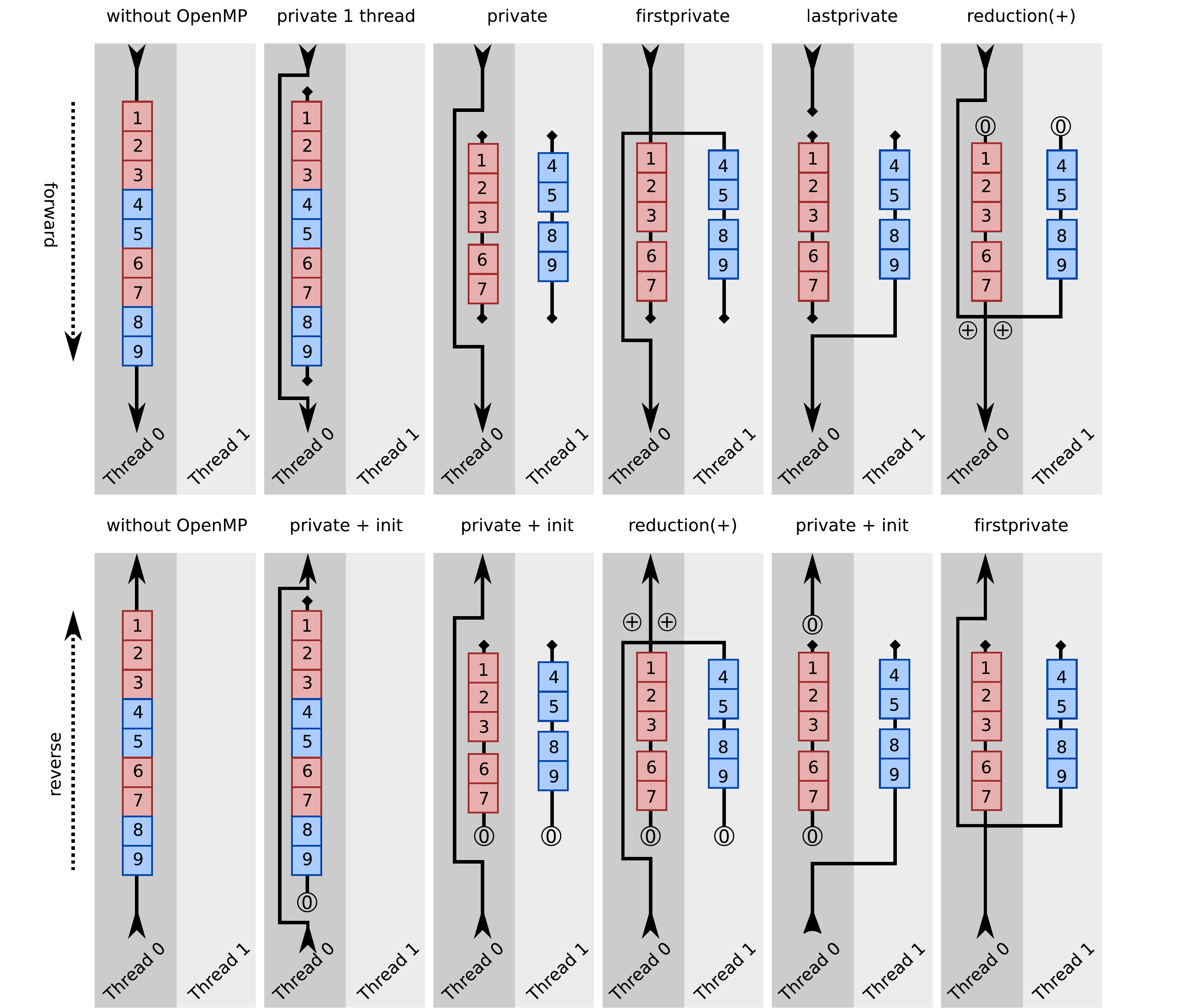}
    \caption{Dataflow of privatized variables using the various OpenMP scoping options. The example shows a loop with 9 iterations (shown as boxes containing the iteration number), which is parallelized by using two threads and a hypothetical schedule that assigns two chunks of iterations to each thread. The iterations are colored based on the thread that they are assigned to in the multithreaded case. The arrows show the dataflow, whereas a break in the arrow shows a variable being newly declared or dropping out of scope (small solid marker), being initialized to zero (marker containing $0$), or being added together (marker containing $+$). The top part shows the behavior for the primal and tangent-mode program; the bottom part shows the corresponding adjoint mode.}
    \label{fig:privateflow}
\end{figure}

Let us first study the case of privatized variables, whose possible dataflows are summarized by Figure~\ref{fig:privateflow}. We start with pure {\tt private} case, whose adjoint is validated by Figure~\ref{fig:privateadj}.
Notice the code introduced in the third section, as the adjoint of ${\tt v}_T$ falling out of scope when the parallel region ends.
Figure~\ref{fig:privateadj} shows that the desired backward-sweep adjoint behavior can be obtained by declaring both {\tt v} and $\overline{\tt v}$ as {\tt private}.
\begin{figure}
\begin{center}
{\small
\begin{tabular}{|l|}
     \hline
     $\hbox{\tt \$OMP PARALLEL PRIVATE(v)}\>\> \{ $\\
     $\>\>\>\>[ {\tt v}=...; \>|\> =...{\tt v}; ]*\> \} $\\
     \hline
     $\forall\>\hbox{\em thread}\>T\>\> \{ $\\
     $\>\>\>\>{\tt v}_T=\_;\>\> [ {\tt v}_T=...; \>|\> =...{\tt v}_T; ]*\>\>\>{\tt v}\!\!\!/_T\> \} $\\
     \hline
     $\forall\>\hbox{\em thread}\>T\>\> \{ $\\
     $\>\>\>\>\hbox{\tt POP}({\tt v}_T);\>\>\overline{\tt v}_T=0;\>\> [ \hbox{\tt POP}({\tt v}_T);\>\> =...\overline{\tt v}_T;\>\> \overline{\tt v}_T=0; \>|\> \overline{\tt v}_T{\tt +\!\!= }...;\>\> =...{\tt v}_T; ]*\>\>\>\>\overline{\tt v}\!\!\!/_T\>\>\>{\tt v}\!\!\!/_T\> \} $\\
     \hline
     $\hbox{\tt \$OMP PARALLEL PRIVATE($\overline{\tt v}$,v)}\>\> \{ $\\
     $\>\>\>\>\hbox{\tt POP}({\tt v});\>\>\overline{\tt v}=0;\>\> [ \hbox{\tt POP}({\tt v});\>\> =...\overline{\tt v};\>\> \overline{\tt v}=0; \>|\> \overline{\tt v}{\tt +\!\!= }...;\>\> =...{\tt v}; ]*\> \} $\\
     \hline
\end{tabular}
}
\end{center}
\caption{Validation diagram for adjoint AD of the \texttt{private} scoping. This diagram as well as the following ones deals only with the backward sweep of the adjoint code. Therefore the stack {\tt PUSH} operations do not appear since they belong to the forward sweep. The adjoint transformation results from the patterns of Figure \ref{fig:elementaryadj}}
\label{fig:privateadj}
\end{figure}

Figure~\ref{fig:firstprivateandreductionadj} shows jointly the case of a {\tt firstprivate} {\tt v}, where all threads initialize their {\tt v}$_T$ with the value of {\tt v} before the parallel region (left diagram), and the symmetric case of a {\tt reduction(+:v)} (right diagram). Since the adjoint of the {\tt firstprivate} initialization is an increment, we obtain a \texttt{reduction}. Conversely the {\tt reduction(+:v)} leads to a \texttt{ firstprivate} $\overline{\tt v}$. We only deal with sum reduction, hence the initialization of {\tt v}$_T$ to zero. Transitions from the third to the fourth sections become clear by comparing with the top two sections of the opposite case.
\begin{figure}
\begin{center}
{\small
\begin{tabular}{|l|}
     \hline
     $\hbox{\tt \$OMP PARALLEL FIRSTPRIVATE(v)}\>\> \{ $\\
     $\>\>\>\>[ {\tt v}=...; \>|\> =...{\tt v}; ]*\> \} $\\
     \hline
     $\forall\>\hbox{\em thread}\>T\>\> \{ $\\
     $\>\>\>\>{\tt v}_T={\tt v};\>\> [ {\tt v}_T=...; \>|\> =...{\tt v}_T; ]*\>\>\>{\tt v}\!\!\!/_T\> \} $\\
     \hline
     $\forall\>\hbox{\em thread}\>T\>\> \{ $\\
     $\>\>\>\>\hbox{\tt POP}({\tt v}_T);\>\>\overline{\tt v}_T=0; $\\
     $\>\>\>\>[\hbox{\tt POP}({\tt v}_T);\>\> =...\overline{\tt v}_T;\>\> \overline{\tt v}_T=0; \>|\> \overline{\tt v}_T{\tt +\!\!= }...;\>\> =...{\tt v}_T; ]* $\\
     $\>\>\>\>\overline{\tt v}{\tt +\!\!= }\overline{\tt v}_T;\>\> \overline{\tt v}_T=0;\>\>\>\overline{\tt v}\!\!\!/_T\>\>\>{\tt v}\!\!\!/_T\>  \} $\\
     \hline
     $\hbox{\tt \$OMP PARALLEL REDUCTION(+:$\overline{\tt v}$) PRIVATE(v)}\>\> \{ $\\
     $\>\>\>\>\hbox{\tt POP}({\tt v});\>\> [ \hbox{\tt POP}({\tt v});\>\> =...\overline{\tt v};\>\> \overline{\tt v}=0; \>|\> \overline{\tt v}{\tt +\!\!= }...;\>\> =...{\tt v}; ]*\> \} $\\
     \hline
\end{tabular}
\hfill
\begin{tabular}{|l|}
     \hline
     $\hbox{\tt \$OMP PARALLEL REDUCTION(+:v)}\>\> \{ $\\
     $\>\>\>\>[ {\tt v}=...; \>|\> =...{\tt v}; ]*\> \} $\\
     \hline
     $\forall\>\hbox{\em thread}\>T\>\> \{ $\\
     $\>\>\>\>{\tt v}_T=0;\>\> [ {\tt v}_T=...; \>|\> =...{\tt v}_T; ]*\>\> {\tt v}{\tt +\!\!= }{\tt v}_T;\>\>\>{\tt v}\!\!\!/_T \} $\\
     \hline
     $\forall\>\hbox{\em thread}\>T\>\> \{ $\\
     $\>\>\>\>\hbox{\tt POP}({\tt v}_T);\>\>\overline{\tt v}_T=0;\>\>\overline{\tt v}_T{\tt +\!\!= }\overline{\tt v}; $\\
     $\>\>\>\>[\hbox{\tt POP}({\tt v}_T);\>\> =...\overline{\tt v}_T;\>\> \overline{\tt v}_T=0; \>|\> \overline{\tt v}_T{\tt +\!\!= }...;\>\> =...{\tt v}_T; ]*\> $\\
     $\>\>\>\>\overline{\tt v}_T=0;\>\>\overline{\tt v}\!\!\!/_T\>\>\>{\tt v}\!\!\!/_T\> \} $\\
     \hline
     $\hbox{\tt \$OMP PARALLEL FIRSTPRIVATE($\overline{\tt v}$) PRIVATE(v)}\>\> \{ $\\
     $\>\>\>\>\hbox{\tt POP}({\tt v});\>\> [ \hbox{\tt POP}({\tt v});\>\> =...\overline{\tt v};\>\> \overline{\tt v}=0; \>|\> \overline{\tt v}{\tt +\!\!= }...;\>\> =...{\tt v}; ]*\> \} $\\
     \hline
\end{tabular}
}
\end{center}
\caption{Validation diagrams for adjoint AD of the \texttt{firstprivate} (left) and \texttt{reduction(+)} (right) scopings. Notice the appearance of the global {\tt v} and $\overline{\tt v}$ in the second and third sections, permitted only by the PGAS notation, and used to expose the implementation of the internal spread or reduction operations. The vector$\times$matrix nature of adjoint AD (Section ~\ref{sec:stsAD}) implies classically that the adjoint AD of ${\tt v}{\tt +\!\!= }{\tt v}_T$ is exactly $\overline{\tt v}_T{\tt +\!\!= }\overline{\tt v}$. We find that adjoint AD makes the \texttt{firstprivate} and \texttt{reduction(+)} scopings commute.}
\label{fig:firstprivateandreductionadj}
\end{figure}

With Figure~\ref{fig:lastprivateadj}, we terminate this review of the {\tt private}-like clauses with the {\tt lastprivate} clause. The OpenMP standard states that the value of a {\tt lastprivate} {\tt v} after the parallel loop is the value of {\tt v}$_T$ at the end of the iteration that would be last in a sequential execution. Implementing this requires an ad hoc predicate {\tt lastiter}.
\begin{figure}
\begin{center}
{\small
\begin{tabular}{|l|}
     \hline
     $\hbox{\tt \$OMP PARALLEL LASTPRIVATE(v)}\>\> \{ $\\
     $\>\>\>\> [ {\tt v}=...; \>|\> =...{\tt v}; ]*\>\> \} $\\
     \hline
     $\forall\>\hbox{\em thread}\>T\>\> \{ $\\
     $\>\>\>\> {\tt v}_T=\_;\>\> [ {\tt v}_T=...; \>|\> =...{\tt v}_T; ]*\>\> \hbox{\tt if (lastiter) v = v}_T;\>\>\>{\tt v}\!\!\!/_T\> \} $\\
     \hline
     $\forall\>\hbox{\em thread}\>T\>\> \{ $\\
     $\>\>\>\> \hbox{\tt POP}({\tt v}_T);\>\> \overline{\tt v}_T=0;\>\> \hbox{\tt if(lastiter)}\> \{\overline{\tt v}_T{\tt +\!\!= }\overline{\tt v};\>\>\overline{\tt v}=0;\}\>\>[ \hbox{\tt POP}({\tt v}_T);\>\> =...\overline{\tt v}_T;\>\> \overline{\tt v}_T=0; \>|\> \overline{\tt v}_T{\tt +\!\!= }...;\>\> =...{\tt v}_T; ]*\>\>\>\overline{\tt v}\!\!\!/_T\>\>\>{\tt v}\!\!\!/_T\>\> \} $\\
     \hline
     $\hbox{\tt \$OMP PARALLEL FIRSTPRIVATE($\overline{\tt v}$) PRIVATE(v)}\>\> \{ $\\
     $\>\>\>\> \hbox{\tt POP}({\tt v});\>\> \hbox{\tt if(!lastiter)}\> \overline{\tt v}=0 ;\>\>[ \hbox{\tt POP}({\tt v});\>\> =...\overline{\tt v};\>\> \overline{\tt v}=0; \>|\> \overline{\tt v}{\tt +\!\!= }...;\>\> =...{\tt v}; ]*\>\> \} $\\
     \noindent $\overline{\tt v}=0; $\\
     \hline
\end{tabular}
}
\end{center}
\caption{Validation diagram for adjoint AD of the \texttt{lastprivate} scoping. A predicate {\tt lastiter} is required to identify the iteration that will provide the exit value of {\tt v}. The PGAS notation lets us refer to global {\tt v} and $\overline{\tt v}$ in the second and third sections to expose the implementation of the internal spread operation. However OpenMP forbids explicit access to global $\overline{\tt v}$ inside the parallel region so, after observing that the piece of PGAS code $\overline{\tt v}_T=0;\>\> \hbox{\tt if(lastiter)}\> \{\overline{\tt v}_T{\tt +\!\!= }\overline{\tt v};\>\>\overline{\tt v}=0;\}$ can be rewritten $\overline{\tt v}_T{\tt =}\overline{\tt v};\>\> \hbox{\tt if(!lastiter)}\>\{\overline{\tt v}_T{\tt =}0;\}\>\> \overline{\tt v}=0;$, we move the final zero-initialization after the region}
\label{fig:lastprivateadj}
\end{figure}

Let us now examine {\tt shared} variables. As we mentioned for tangent differentiation, only three acceptable cases must be examined: {\tt v} is accessed in any manner but exclusively by one thread, or {\tt v} is only read but by any thread, or {\tt v} is only incremented by any thread. In this last case, the code is deterministic and therefore acceptable only if all increments are declared \texttt{atomic}. Figure~\ref{fig:exclusiveandincrementadj} shows the validation diagrams for the exclusive single-thread access and the atomic-increment-only cases.
For both, {\tt v} and $\overline{\tt v}$ may be shared in the adjoint parallel region.
\begin{figure}
\begin{center}
{\small
\begin{tabular}{|l|}
     \hline
     $\hbox{\tt \$OMP PARALLEL SHARED(v)}\>\> \{ $\\
     $\>\>\>\> [{\tt v}=...; \>|\> =...{\tt v}; ]*\> \} $\\
     \hline
     $\hbox{\em On only one thread}\>T\>\> \{ $\\
     $\>\>\>\> [{\tt v}=...; \>|\> =...{\tt v}; ]*\> \} $\\
     \hline
     $\hbox{\em On only one thread}\>T\>\> \{ $\\
     $\>\>\>\> [ \hbox{\tt POP}({\tt v});\>\> ...=\overline{\tt v};\>\> \overline{\tt v}=0; \>|\> \overline{\tt v}{\tt +\!\!= }... ;\>\> =...{\tt v}; ]*\> \} $\\
     \hline
     $\hbox{\tt \$OMP PARALLEL SHARED($\overline{\tt v}$,v)}\>\> \{ $\\
     $\>\>\>\> [ \hbox{\tt POP}({\tt v});\>\> ...=\overline{\tt v};\>\> \overline{\tt v}=0; \>|\> \overline{\tt v}{\tt +\!\!=}... ;\>\> =...{\tt v}; ]*\> \} $\\
     \hline
\end{tabular}
\hfill
\begin{tabular}{|l|}
     \hline
     $\hbox{\tt \$OMP PARALLEL SHARED(v)}\>\> \{ $\\
     $\>\>\>\>[\> {\tt v}\hbox{\framebox {\makebox[\totalheight]{\tt +=}}}...; ]*\> \} $\\
     \hline
     $\forall\>\hbox{\em thread}\>T\>\> \{ $\\
     $\>\>\>\>[\> {\tt v}\hbox{\framebox {\makebox[\totalheight]{\tt +=}}}...; ]*\> \} $\\
     \hline
     $\forall\>\hbox{\em thread}\>T\>\> \{ $\\
     $\>\>\>\>[\> =...\overline{\tt v}; ]*\> \} $\\
     \noindent $\hbox{\tt POP}({\tt v}); $\\
     \hline
     $\hbox{\tt \$OMP PARALLEL SHARED($\overline{\tt v}$,v)}\>\> \{ $\\
     $\>\>\>\>[\> =...\overline{\tt v}; ]*\> \} $\\
     \noindent $\hbox{\tt POP}({\tt v}); $\\
     \hline
\end{tabular}
}
\end{center}
\caption{Validation diagrams for the adjoint of a \texttt{shared} reference, in the exclusive single-thread access case (left) and in the atomic-increment-only case (right). These are the favorable cases where the scoping of {\tt v} can be extended to $\overline{\tt v}$.
On the right, the $\hbox{\tt POP}({\tt v})$ needed in general for the adjoint of an increment of {\tt v} (see Figure~\ref{fig:elementaryadj}) is moved after the parallel region, and the corresponding $\hbox{\tt PUSH}({\tt v})$, not shown here, is moved accordingly. This is necessary to keep {\tt v} shared as concurrent overwrites are forbidden. This is permitted because by hypothesis the primal parallel region contains no read nor overwrite of {\tt v}.}
\label{fig:exclusiveandincrementadj}
\end{figure}

\begin{figure}
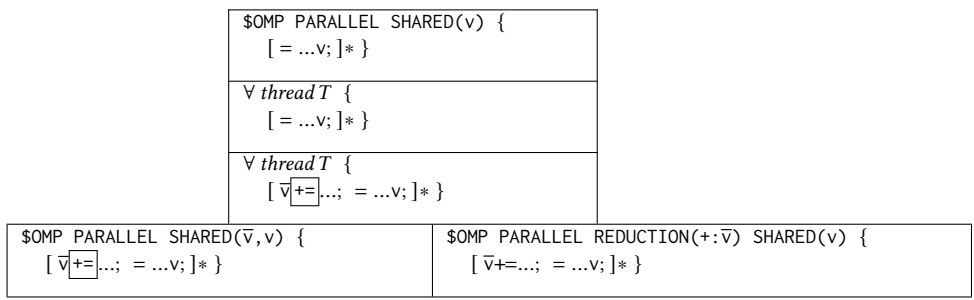

\begin{center}
{\small
\begin{tabular}{llll}
     \cline{2-3}
     \phantom{MMMMMMMMMMM} & \multicolumn{2}{|l|}{
     \begin{minipage}[t]{.3\linewidth}
     $\hbox{\tt \$OMP PARALLEL SHARED(v)}\>\> \{ $\\
     \phantom{M} $[\> =...{\tt v}; ]*\> \} $\\
     \end{minipage}
     } &\\
     \cline{2-3}
     \phantom{MMMMMMMMMMM} & \multicolumn{2}{|l|}{
     \begin{minipage}[t]{.3\linewidth}
     $\forall\>\hbox{\em thread}\>T\>\> \{ $\\
     \phantom{M} $[\> =...{\tt v}; ]*\> \} $\\
     \end{minipage}
     } &\\
     \cline{2-3}
     \phantom{MMMMMMMMMMM} & \multicolumn{2}{|l|}{
     \begin{minipage}[t]{.3\linewidth}
     $\forall\>\hbox{\em thread}\>T\>\> \{ $\\
     \phantom{M} $[\> \overline{\tt v}\hbox{\framebox {\makebox[\totalheight]{\tt +=}}}...;\>\> =...{\tt v}; ]*\> \} $\\
     \end{minipage}
     } &\\
     \hline
     \multicolumn{2}{|l|}{
     \begin{minipage}[t]{.35\linewidth}
     $\hbox{\tt \$OMP PARALLEL SHARED($\overline{\tt v}$,v)}\>\> \{ $\\
     \phantom{M} $[\> \overline{\tt v}\hbox{\framebox {\makebox[\totalheight]{\tt +=}}}...;\>\> =...{\tt v}; ]*\> \} $\\
     \end{minipage}
     } &
     \multicolumn{2}{l|}{
     \begin{minipage}[t]{.45\linewidth}
     $\hbox{\tt \$OMP PARALLEL REDUCTION(+:$\overline{\tt v}$) SHARED(v)}\>\> \{ $\\
     \phantom{M} $[\> \overline{\tt v}{\tt +\!\!= }...;\>\> =...{\tt v}; ]*\> \} $\\
     \end{minipage}
     } \\
     \hline
\end{tabular}
}
\end{center}
\caption{Validation diagram for the adjoint of a \texttt{shared} reference, in the read-only case.
The desired PGAS behavior of the adjoint can be achieved by two alternative OpenMP implementations, unfortunately none of them simply extending the \texttt{shared} scoping to $\overline{\tt v}$.}
\label{fig:atomicreadreadadj}
\end{figure}

Figure~\ref{fig:atomicreadreadadj} shows the more delicate case of read-only access.
It shows two possible answers.
On the left is a deceivingly simple answer that also declares a shared $\overline{\tt v}$. Unfortunately it requires an {\tt atomic} clause on each increment of $\overline{\tt v}$, which may be costly. For this reason we might prefer the option on the right, which achieves an equivalent behavior with a {\tt reduction(+)} clause. This takes care of the access conflicts, since $\overline{\tt v}$ is indeed expanded as a private $\overline{\tt v}_T$. On the other hand this option will create copies of {\tt v} for each thread, and memory for each thread may be limited.

\begin{figure}
    \centering
    \includegraphics[width=0.9\linewidth]{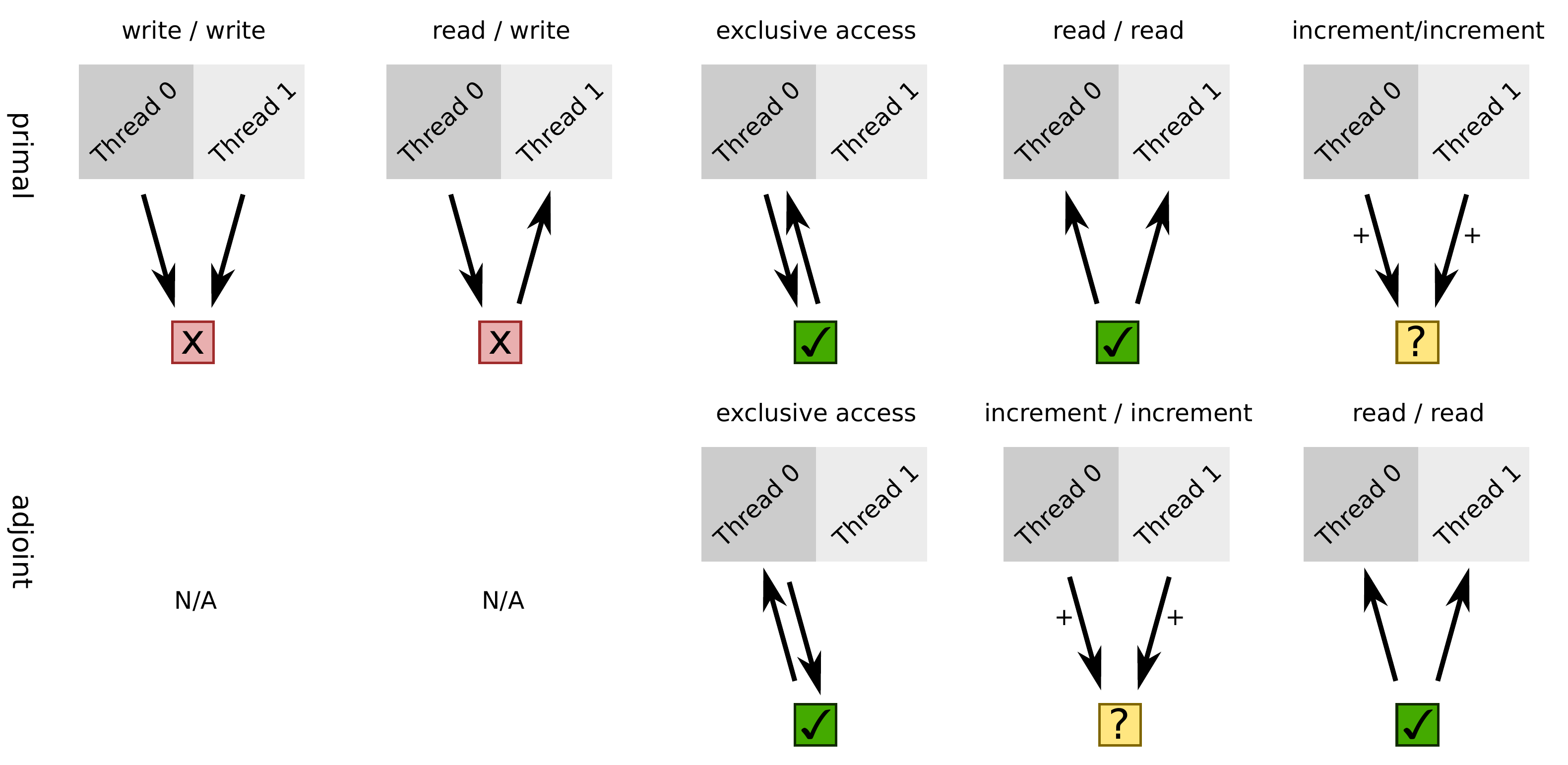}
    \caption{Various ways in which a primal parallel program can access a shared-memory location (top row) and what the corresponding adjoint program must do to ensure correct differentiation (bottom row). The first two cases show a write/write or read/write conflict, which would cause undefined behavior. We assume that our input programs do not contain these, although in general checking this statically is impossible. Therefore, if a primal program contains write access (possibly in addition to read access) to a shared-memory location, no other thread can access that same data within the same parallel region, and differentiation can assume exclusive access. The biggest concern is the existence of concurrent read access in the primal, since this leads to concurrent increments, which must be safeguarded by use of atomics or reductions in the adjoint. Parallel increment/increment access, if safeguarded with atomics, is also legal and leads to concurrent read access in the adjoint, which is safe.}
    \label{fig:sharedvars}
\end{figure}

Figure~\ref{fig:sharedvars} summarizes this study of scoping for the adjoint of a shared variable. Since we assumed that the primal code is correct and deterministic, we have only three cases, but these cases call for possibly different scopings. The problem now is that an array may have cells resorting to the cases of Figure~\ref{fig:exclusiveandincrementadj} and others to the read-only case of Figure~\ref{fig:atomicreadreadadj}. OpenMP does not let us declare scopings cell per cell, and even that would not help because even a scalar variable may resort to several cases, depending on runtime or on undecidable properties. To begin with, our fallback option to exploit the parallel properties that the adjoint region inherits from the primal region is to declare $\overline{\tt v}$ as {\tt shared} and label as {\tt atomic} every increment access to $\overline{\tt v}$, in other words, to ignore the \texttt{reduction(+)} alternative of Figure~\ref{fig:atomicreadreadadj}.

Fortunately, we can prove that this fallback needs to declare \texttt{atomic} only the increments to $\overline{\tt v}$ and not the isolated write or read accesses to $\overline{\tt v}$: in Figures~\ref{fig:exclusiveandincrementadj} and~\ref{fig:atomicreadreadadj}, we observe that plain write or read accesses to $\overline{\tt v}$ occur only in Figure~\ref{fig:exclusiveandincrementadj}. In other words if one particular reference $\overline{\tt v}$ is plainly read or written, then the corresponding primal reference {\tt v} is accessed either by only one thread or by only atomic increments. Since we assume that the primal program is deterministic, this forbids that this primal memory cell is read by more than one thread. Therefore there can be no increment access conflict on the $\overline{\tt v}$ memory cell.

Still, this fallback suffers from the cost of the {\tt atomic} pragma, and it is profitable to detect when {\tt v} resorts never, or only, to the read/read case. Then the AD tool can generate a more efficient adjoint code. Some static analyses for detection have been proposed~\cite{forster2014algorithmic} but cannot work for general programs, particularly if the memory access pattern depends on runtime input. Algorithm~\ref{fig:decisionTree} shows a possible decision tree for any {\tt shared} primal variable {\tt V}, informed by detection results. In addition, since detection is a static and therefore approximate analysis, we will allow the user to override the tool's choices with a pragma.

\begin{algorithm}
  \uIf{(for all cell {\tt v} of {\tt V}, all accesses to {\tt v} in the primal parallel region are (atomic) increments \normalfont{\textbf{or}}\\\phantom{if for all cell {\tt v} of {\tt V}, (} {\it all accesses to {\tt v} in the primal parallel region occur in the same thread)}}{
    declare {\tt SHARED($\overline{\tt V}$)}
  }
  \uElseIf{(all accesses to {\tt V} in the primal parallel region are read accesses \normalfont{\textbf{and}}\\\phantom{else if..(} {\tt V} is not too big to be privatized)}{
    declare {\tt REDUCTION(+:$\overline{\tt V}$)}
  }
  \Else{
    declare {\tt SHARED($\overline{\tt V}$)} and declare {\tt atomic} all increments of $\overline{\tt V}$
  }
 \caption{Rudimentary decision tree to choose the scoping of the adjoint $\overline{\tt V}$ of a variable {\tt V}, based on the access patterns detected for each memory cell {\tt v} of {\tt V}. Trade-off between reduction and atomic increments should be elaborated further, considering the ``density'' of conflicts and updates, as well as array size and characteristics of the platform. See Section~\ref{sec:testcases} for examples.}
\label{fig:decisionTree}
\end{algorithm}

\section{Implementation}
\label{sec:implementation}

The Tapenade AD tool~\footnote{https://gitlab.inria.fr/tapenade/tapenade/} had to be extended in various ways to support OpenMP, touching the parser, code analysis, and code generation stages. These extensions are discussed in Section~\ref{sec:tapenadeinternal}.
The generated code interacts with a runtime support library that is part of the Tapenade distribution and facilitates the differentiation of OpenMP \emph{schedules}, explained in Section~\ref{sec:schedulereversal}, and provides a thread-safe stack mechanism to store and retrieve intermediate values, shown in Section~\ref{sec:stack}\footnote{https://gitlab.inria.fr/tapenade/tapenade/-/tree/develop/openmp/OMP\_FirstAidKit}.

\subsection{Changes to the Tapenade AD Tool}
\label{sec:tapenadeinternal}
Following the workflow of the AD tool, we first extended the parser to analyze OpenMP syntax. The only point of note here is that we defined a command line option to turn the OpenMP extension on and off, allowing for a parsing mode that oversees OpenMP pragmas as plain comments. This is needed for backward compatibility and, as we mentioned, because turning on OpenMP support may change the dataflow for {\tt private} variables. Many OpenMP compilers have a similar option. To date, we have extended only our Fortran parser. However, all the changes that we have done farther down the workflow apply independently of the source language, so that it will only take adaption of our C parser to obtain the same differentiation capacity on C. We then extended correspondingly the abstract syntax that Tapenade takes as input, adding tree constructs such as {\tt parallelRegion} and {\tt parallelLoop} and constructs for scoping and scheduling clauses. These constructs are naturally structured; for example, a {\tt parallelRegion} or {\tt parallelLoop} both have two children trees: one for the list of clauses and the other for the complete code contained.

Tapenade then builds its internal representation. Similarly to other control structures, a {\tt parallelRegion} or a {\tt parallelLoop} gives birth to a basic block in the control-flow graph that controls entry into and exit from the subflow graph of the contained region or loop. This basic block just contains the ``header'' of the control structure, that is, its name and clauses. This almost unchanged structure for the control-flow graph allows all dataflow analysis to run with virtually no modification. Since we saw in Section~\ref{sec:modelVars} that the scoping of differentiated variables depends on the scoping of the primal, we have added a simple analysis to propagate scoping from declaration location in clauses to locations where variables are used.

\begin{figure}
    \centering
    \includegraphics[width=0.8\linewidth]{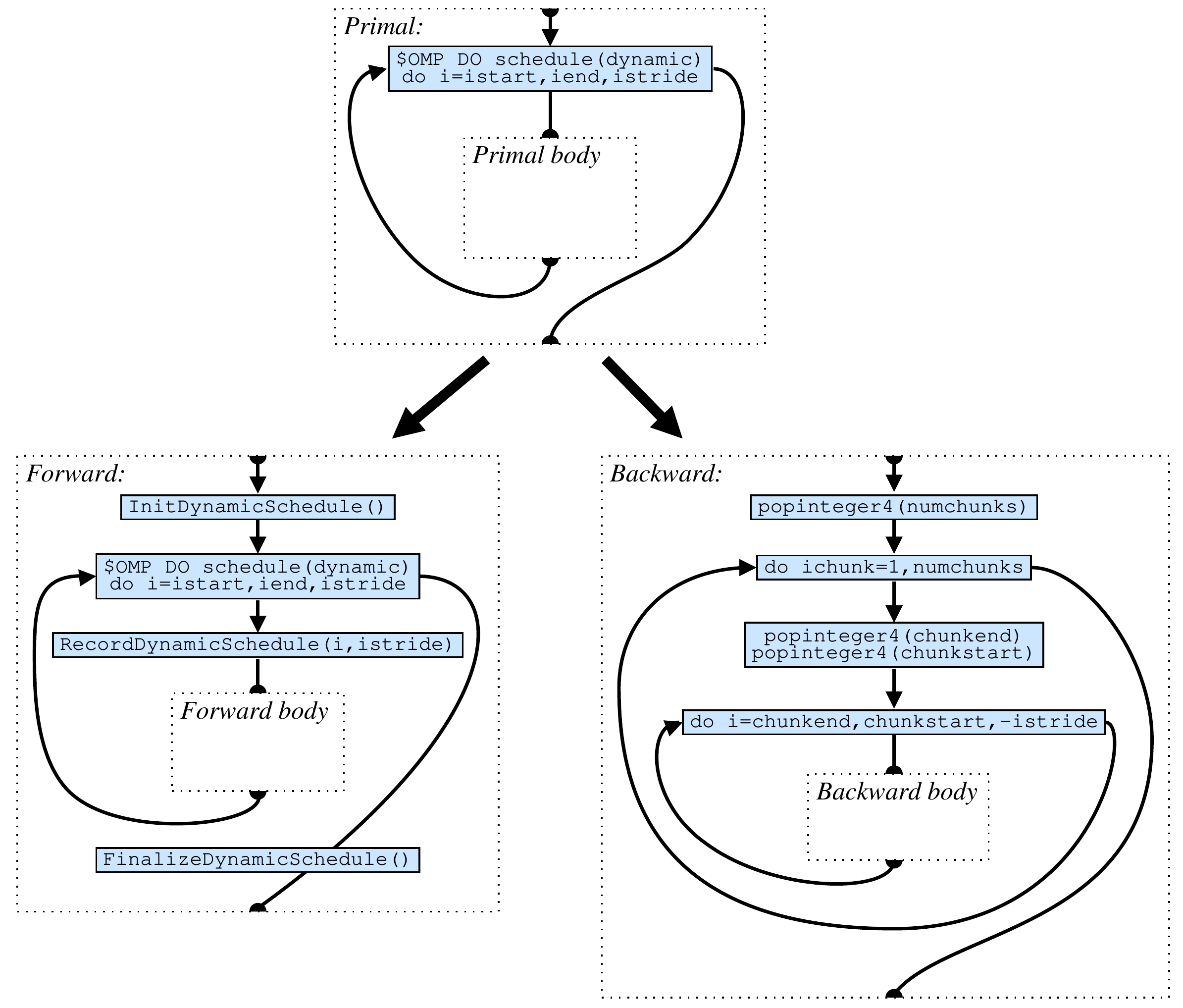}
    \caption{Adjoint flow graph (with dynamic scheduling option) of a primal parallel {\tt \$OMP DO} loop. Assuming the {\it Primal body} is recursively differentiated as a pair of {\it Forward body} and {\it Backward body} for the forward and backward sweeps, the figure shows the pair of flow graphs built for the adjoint differentiation of the parallel {\tt \$OMP DO}}.
    \label{fig:OmpDoDynamic}
\end{figure}

The most significant development is about differentiation, in the tool part that deals with building the differentiated flow graph. At that stage, the flow graph appears as a tree of nested flow graphs, so that the transformation to apply to, for example, a \texttt{parallelLoop} applies regardless of the transformation recursively applied to the inner or outer levels. Without going into implementation details irrelevant here, we show in Figure~\ref{fig:OmpDoDynamic} the flow graph level that the AD tool builds as the adjoint of a primal \texttt{parallelLoop} level of the flow graph, when dynamic scheduling is selected (see  Section~\ref{sec:schedulereversal}).
The remaining changes deal with the variable scoping of the differentiated variables. The AD tool creates this new scoping following the rules described in Section~\ref{sec:modelVars}, then adds it to the clauses of the generated OpenMP pragmas for tangent or backward-sweep adjoint flow graphs. When it cannot be avoided, increments of {\tt shared} adjoint variables are declared {\tt atomic}. The OpenMP standard disallows atomic updates to variables with the \texttt{allocatable} attribute, which would instead require updates to be made inside a \texttt{critical} section. Similarly, atomic updates can only be made to scalar variables (or a single entry in an array), whereas updates to slices of arrays need to be written as a sequence of atomic scalar updates, or wrapped into a \texttt{critical} section. Our current implementation does not detect such cases, but we anticipate that this deficiency will be straightforward to remove in future versions.

As a last resort, Tapenade provides a differentiation directive ({\tt \$ad omp}) that the user can place in the primal code to override the scopings of differentiated variables in the tangent or in the adjoint code. The syntax is similar to that of OpenMP pragmas, except that it introduces a new clause \texttt{atomic} meaning actually \texttt{shared} with all increments \texttt{atomic}. For example,
\begin{lstlisting}
!$ad omp_adjoint shared(a) reduction(+:b) atomic(c)
!$omp parallel do shared(a,b,c)
\end{lstlisting}
forces Tapenade to declare, in the adjoint code, scoping \texttt{reduction(+)} for variables $\overline{\tt b}$, \texttt{shared} for $\overline{\tt a}$ and $\overline{\tt c}$, and to label all increments to $\overline{\tt c}$ as \texttt{atomic}.

\subsection{Thread-Safe Stack}
\label{sec:stack}

The adjoint mode of AD commonly requires intermediate values to be stored during the forward sweep, which can then be retrieved during the backward sweep. The conventional approach for sequential programs is to use a stack, since the intermediate values are typically needed in the exact reverse order in which they were stored, making a last-in-first-out data structure a perfect fit.
For OpenMP-parallel programs, the order in which intermediate values are produced is still deterministic within a given thread, but different threads may run at different speeds relative to each other. We thus give each thread its own \emph{threadprivate} stack. To ensure that the values are available on the correct thread and on top of the stack at the correct point in the backward sweep, we therefore also need to ensure that each thread will perform all reverse loop iterations that correspond to the forward iterations that were performed by that same thread. These features are discussed in more detail in Section~\ref{sec:schedulereversal}. Alternatively, one could create a separate stack for each loop index, which would make the stack mechanism immune to changes in the loop schedule between the forward and reverse sweep. On the other hand, the large number of stacks could create a significant overhead (particularly if there are many iterations that each only store one or few values on their stack). Correct schedule reversal as discussed in the upcoming section would still be required to capture the correct data flow on each thread, and to avoid NUMA-related performance issues that could arise if a stack is created and consumed on different threads.  We therefore have not explored this option further.

We evaluated two options of implementing the threadprivate stack. The first option is to use the STL stack implementation from the C++ standard library. Since Tapenade supports source transformation for programs written in C and Fortran, this option further required a C/Fortran wrapper to make the STL stack accessible. A separate stack object is created for each data type (float, double, int, char, pointers to various types) that may need to be stored by the differentiated program to capture intermediate values or control flow decisions or array indices that may affect the derivative computation. The stack objects are declared as threadprivate using the OpenMP \texttt{threadprivate} pragma.

The second option, which was ultimately chosen for the production version, is a customized stack written in pure C. The stack consists of a bidirectional linked list of blocks that hold data. A very small block size harms the performance of the stack, since new blocks need to be dynamically allocated and some pointers updated whenever a block is filled up. On the other hand, a very large block size can be wasteful, since more memory might be allocated on each thread than is required. We chose a block size of $64$ kilobytes, since we found no significant performance improvements for block sizes beyond this. Each thread holds its own threadprivate pointer to the start and end of this linked list, as well as a counter to hold the fill status of the current block. Threads allocate additional blocks to their own stack when needed.

Values are stored in the same stack regardless of their data type, and values take only the amount of space needed for that type. One value might be split up over two blocks if the current block has fewer bytes available than required. For example, an 8-byte double-precision value might have its first 3 bytes stored at the end of one block and the remaining 5 bytes at the start of the subsequent block.

The custom-made stack has additional functionality compared with a conventional last-in-first-out stack. It is sometimes necessary to execute multiple backward sweeps for a particular function, for example in computing adjoints of fixed-point iterative methods~\cite{Taftaf2014AoF}. To accommodate this, the stack can store its current top pointer, then pop a range of values without deleting them, and subsequently restore its top pointer to the stored position.

\subsection{Schedule Reversal}
\label{sec:schedulereversal}

%As mentioned previously,
The reverse mode must be implemented so that each thread computes the derivatives of all operations that were executed on that same thread, to ensure the correct dataflow through privatized variables and to ensure that intermediate results are available on the correct threadprivate stack.

As a result, for parallel loops the same number of iterations must be performed on each thread in both the forward and backward sweeps. Furthermore, the loop counter often affects the actions of the loop body, for example through its use in array index expressions or branch conditions; and therefore each reverse iteration must recompute or restore the loop counter value of the corresponding forward iteration. The sequence of values that the loop counter assumes in the forward sweep must be exactly reversed in the backward sweep on the same thread.

This problem is closely related to the OpenMP \emph{scheduling}, that is, the mapping of iterations to threads. In general, OpenMP schedules are not straightforward to reverse. The schedule types \texttt{guided}, \texttt{auto}, and \texttt{dynamic} are nondeterministic; and a reversal thus always requires runtime recording, paired with a customized scheduler for the backward sweep that operates on this recording.
Even the \texttt{static} schedule, despite being deterministic, maps iterations to threads in such a way that a reversed loop counter sequence cannot be achieved through an OpenMP \texttt{static} schedule, at least for loops whose number of iterations is not a multiple of the number of threads.

We developed two solutions to reverse OpenMP: static and dynamic schedules. Both solutions work by replacing an OpenMP parallel worksharing loop (in which the schedule is determined by OpenMP) with a parallel region containing a normal (non-OpenMP) loop with loop bounds that are determined by using our custom, AD-specific runtime routines. Because of this transformation, any clauses associated with the worksharing loop must be moved to the parallel region, which does not change the semantics except in the case of the  \texttt{lastprivate} clause, which can only be associated with a parallel loop, not a parallel region. Tapenade solves this by assigning a \texttt{shared} scope to the original variable, and introducing an auxiliary \texttt{private} variable that is used instead of the lastprivate variable within the loop. After the loop, the thread that performed the work corresponding to the final logical iteration in the original iteration space must copy the value of its auxiliary variable into the shared original variable.

\subsubsection{Static Schedules}
According to the OpenMP standard, when a chunk size is explicitly specified, the mapping of iterations to threads in a statically scheduled work-sharing loop is clearly defined. But the OpenMP standard does not specify the exact mapping of iterations to threads when no chunk size is specified, making it impossible to reverse the OpenMP-provided schedule without recording or relying on implementation-defined behavior.

Even a schedule with a fixed chunk size is not easy to revert if the number of threads does not evenly divide the number of iterations, since the last chunk of the forward sweep will be smaller (containing only a remainder). Consequently, a correct schedule reversal must place the corresponding reverse iterations into a smaller chunk, but those iterations are the sequentially first iterations in the backward sweep and will be incorrectly placed into a normal-sized chunk by the OpenMP scheduler.

We note that there are simple cases where an exact schedule reversal is not necessary, for example during a parallel summation of two arrays into an equally-sized result array, but we do not attempt to take advantage of such situations, and instead always use a custom static schedule that can be reversed easily. This schedule is part of our new Tapenade OpenMP runtime library. OpenMP work-sharing loops that explicitly use a static schedule through the \texttt{schedule(static)} clause are replaced by Tapenade with a parallel region wherein each thread calls the static scheduler. Based on the original loop's iteration space, the number of threads, and the ID of the calling thread, the scheduler returns loop bounds for this thread. In the backward sweep, each thread simply flips the direction of its loop. Below is an example of a parallel loop using the custom-made static scheduler.

\begin{lstlisting}
! original code
!$omp parallel do schedule(static)
do i=istart,iend,istride
  ! ... loop body
end do

! transformed forward sweep
!$omp parallel
call getStaticSchedule(istart,iend,istride,chunkstart,chunkend)
do i=chunkstart,chunkend,istride
  ! ... loop body
end do
!$omp end parallel

! generated backward sweep
!$omp parallel
call getStaticSchedule(istart,iend,istride,chunkstart,chunkend)
do i=chunkend,chunkstart,-istride
  ! ... loop body
end do
!$omp end parallel
\end{lstlisting}

The \texttt{getStaticSchedule} function internally calls OpenMP functions to determine the number of threads and the ID of the calling thread. Since our static scheduler is similar to those shipped with common compilers (and in the forward sweep, functionally equivalent with that provided by LLVM), we would not expect any significant performance difference from using our own schedule. We confirmed this in a simple a parallel array copy experiment, in which we compared the OpenMP static schedule with our own schedule.

\subsubsection{Dynamic Schedules}
The approach for dynamic schedules requires three function calls to be inserted into the forward sweep: an initialization call inside the parallel region just before the start of the worksharing loop, a recording call in each iteration of the loop, and a finalization call after the end of the worksharing loop but still inside the parallel region. The recording call stores the current loop counter value in a threadprivate variable for the next iteration. In each loop iteration, the recording function compares the current and previous loop counter values. Within a chunk, the difference between the current and previous counter must be identical to the loop stride, and any other difference indicates that the previous iteration must have been the last iteration of a previous chunk, while the current iteration must be the first iteration of a new chunk. When this situation happens,  the previous chunk end and current chunk start are stored on the threadprivate AD stack, and a threadprivate counter for the number of chunks is incremented. While this recording function detects the ``jumps'' between chunks, the initialization and finalization functions are responsible for recording the start of the first chunk and the end of the last chunk, respectively, as well as storing the final number of chunks for each thread. Note that a thread may be assigned two immediately adjacent chunks by the OpenMP run time, in which case the recording function would not detect a jump in the iteration counter, and would thus record multiple small chunks as just one large chunk. This does not affect the correctness of the schedule reversal.

\begin{lstlisting}
! original code
!$omp parallel do schedule(dynamic)
do i=istart,iend,istride
  ! ... loop body
end do

! transformed forward sweep
!$omp parallel private(i)
call initdynamicschedule()
!$omp do schedule(dynamic)
do i=istart,iend,istride
  call recorddynamicschedule(i,istride)
  ! ... forward sweep loop body
end do
call finalizedynamicschedule()
!$omp end parallel

! generated backward sweep
!$omp parallel private(numchunks,ichunk,chunkstart,chunkend,i)
call popinteger4(numchunks)
do ichunk=1,numchunks
  call popinteger4(chunkend)
  call popinteger4(chunkstart)
  do i=chunkend,chunkstart,-istride
    ! ... backward sweep loop body
  end do
end do
!$omp end parallel
\end{lstlisting}

\begin{figure}
\begin{tikzpicture}
\begin{axis}[
ymin=0.008,
ymax=15,
xmin=1,
xmax=65536,
legend style={at={(-1.07,1)},
      anchor=north west,legend columns=4},
xtick={1,16,256,4095,65536},
ytick={0.01,0.1,1,10},
xticklabels={1,16,256,4095,65536},
yticklabels={0.01,0.1,1,10},
xlabel={Chunk Size},
ylabel={Run Time [s]},
title={icc},
title style={font=\bfseries},
xmode=log,
ymode=log,
xmajorgrids,
ymajorgrids,
width=19em,
height=17em,
cycle list name=cbw,
scale only axis
]
\addplot table [x expr=\thisrow{chunk}, y expr=\thisrow{dynamic}] {\schedicctable};% \addlegendentry{dynamic}
\addplot table [x expr=\thisrow{chunk}, y expr=\thisrow{forward}] {\schedicctable};% \addlegendentry{dynamic record}
\addplot table [x expr=\thisrow{chunk}, y expr=\thisrow{reverse}] {\schedicctable};% \addlegendentry{reverse replay}
\end{axis}
\end{tikzpicture}\hspace{-12.7em}
\begin{tikzpicture}
\begin{axis}[
ymin=0.008,
ymax=15,
xmin=1,
xmax=65536,
legend style={at={(-0.8,0)},
      anchor=south west,legend columns=4},
xtick={1,16,256,4095,65536},
ytick={0.01,0.1,1,10},
xticklabels={1,16,256,4095,65536},
yticklabels={0.01,0.1,1,10},
xlabel={Chunk Size},
ylabel={Run Time [s]},
title={gcc},
title style={font=\bfseries},
xmode=log,
ymode=log,
xmajorgrids,
ymajorgrids,
width=19em,
height=17em,
cycle list name=cbw,
scale only axis
]
\addplot table [x expr=\thisrow{chunk}, y expr=\thisrow{dynamic}] {\schedgcctable}; \addlegendentry{dynamic}
\addplot table [x expr=\thisrow{chunk}, y expr=\thisrow{forward}] {\schedgcctable}; \addlegendentry{dynamic record}
\addplot table [x expr=\thisrow{chunk}, y expr=\thisrow{reverse}] {\schedgcctable}; \addlegendentry{reverse replay}
\end{axis}
\end{tikzpicture}
\caption{Time to copy an array of 183.5M single precision values from one array into another in an OpenMP parallel loop using 28 Skylake cores and dynamic scheduling (black), with additional schedule recording by Tapenade (blue), and while replaying a previously recorded dynamic schedule in reverse (red), for a range of chunk sizes. The overhead of OpenMP's dynamic scheduling is so large for small chunk sizes that the cost of schedule recording is insignificant. Replaying a pre-recorded schedule using our run time library is cheaper by one or two orders of magnitude in these cases. For larger chunk sizes, the cost of the dynamic schedule decreases.  \textbf{Left:} With Intel compiler and run time, our schedule recording has a measurable overhead for large chunk sizes. \textbf{Right:} The GNU compiler performs better, and recording has no measurable time cost in this case.}
\label{fig:dyschedoverhead}
\end{figure}
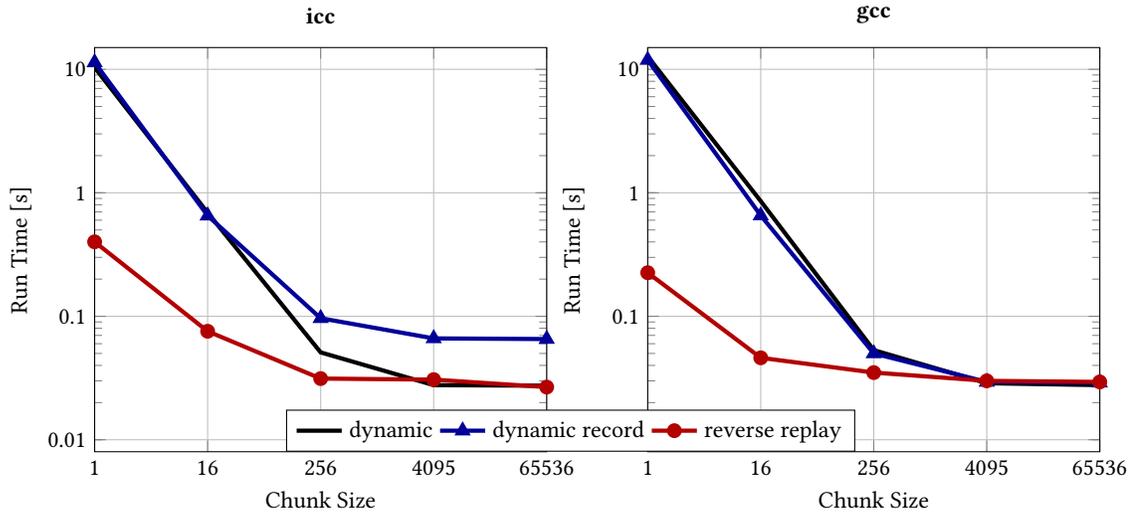

The schedule recording approach works for arbitrary schedules (including the static schedule) and is therefore the default choice in our implementation in Tapenade for parallel loops that do not explicitly set a static schedule through the \texttt{schedule(static)} clause. The flexibility of this approach comes at the cost of branching and assignment operations to detect the start and end of chunks, and the memory footprint for storing two integers (chunk start and end) for each chunk. For loops with many iterations, the user can reduce the memory footprint by increasing the chunk size. There can be a measurable time overhead associated with this, as shown in Figure~\ref{fig:dyschedoverhead}.

\pgfplotstableread{
chunk	dynamic	forward	reverse
1	10.34900808	11.35687113	0.4007019997
16	0.6898082018	0.6526473999	7.56E-02
256	5.10E-02	9.64E-02	3.13E-02
4096	2.77E-02	6.62E-02	3.07E-02
65536	2.74E-02	6.54E-02	2.67E-02
}\schedicctable
\pgfplotstableread{
chunk	dynamic	forward	reverse
1	12.7205801	11.88886724	0.2252777771
16	0.8556366423	0.6544748411	4.61E-02
256	5.32E-02	5.00E-02	3.50E-02
4096	2.87E-02	2.94E-02	3.00E-02
65536	2.77E-02	2.90E-02	2.94E-02
}\schedgcctable

\section{Test Cases}
\label{sec:testcases}

We evaluate our AD implementation in Tapenade by applying it to the scientific computing test cases introduced in the following subsections and measuring the performance of the generated derivative programs on a
multicore system. We compute first-order derivatives. The full source code of all test cases is available in the Tapenade repository\footnote{https://gitlab.inria.fr/tapenade/tapenade/-/tree/develop/openmp/examples}. Measurements were performed on a system with two Intel Xeon Platinum 8180M CPUs (\emph{Skylake}), only one of which is used for our experiments. We use the \texttt{OMP\_PLACES=sockets(1)} environment variable to ensure that all threads are scheduled on the same CPU to avoid NUMA issues and prevent thread migration. The processor has 28 physical cores and supports up to 56 threads through hyperthreading. We use the Intel \texttt{icc} compiler version 19.0.4.243 with the compiler flags \texttt{-qopenmp -O3} to enable OpenMP support and optimization.
For all test cases, we create the following executables:
\begin{description}
    \item[Primal] A program calling the original function
    \item[Tangent] A program calling the function generated by tangent-linear AD
    \item[Adjoint Serial] A program calling the function generated by adjoint AD after removing all OpenMP pragmas, with no shared-memory parallelism
    \item[Adjoint Atomic] A program calling the generated adjoint function wherein concurrent increments to a shared variable are resolved using OpenMP \texttt{atomic} pragmas
    \item[Adjoint Reduction] A program calling the generated adjoint function wherein variables with concurrent increments are declared as an OpenMP \texttt{reduction}
\end{description}
For the primal and tangent programs, we did not observe significant runtime differences between an OpenMP-parallel executable run on one thread and an executable compiled without OpenMP support. We therefore did not create separate serial primal or tangent programs. The privatization that is necessary in the adjoint reduction case, as well as the atomics in the adjoint atomic case, has a significant performance overhead; and the performance of the OpenMP parallel programs executed on one thread is worse than that of the nonparallelized program. The adjoint scaling results in the following sections are obtained by comparing the parallel execution time with the runtime of the cheaper serial executable and not with the runtime of the parallel executable on one thread. Thus the scaling factors look less impressive but are a more honest assessment of the speedup obtained through the new OpenMP support.

Besides scaling, we report absolute runtimes for the serial and parallel programs. The reported serial times are always obtained from the serial program version if it exists, and the parallel time reported is the best observed time for any number of threads tried, since the fastest configuration may be using 56, 28, or in some cases fewer threads on our system. The fastest configuration can usually be determined by finding the maximum speedup in the corresponding scalability plot.

\subsection{Stencil Kernel}
Stencil kernels are a common motif in a wide range of applications including structured-mesh solvers, linear algebra, image processing, and convolutional layers in deep learning. Stencil kernels are characterized by a loop that in each iteration updates indices in an output array based on neighboring indices in an input array, with a simple relationship between loop counter, input array indices, and output array indices. For example, consider this implementation of a \emph{three-point stencil}.
\begin{lstlisting}
do i=2,n-1
  arr_out(i) = a*arr_in(i-1) + b*arr_in(i) + c*arr_in(i+1)
end do
\end{lstlisting}
Stencil kernels can operate on multidimensional arrays, which is often implemented as a deeper loop nest, and each output value may depend on a larger number of neighbors in each direction, for example in a \emph{seventeen-point stencil} \texttt{arr\_in(i-8)}, \texttt{...}, \texttt{arr\_in(i+8)}. We will refer to the one-dimensional three-point and seventeen-point stencil as small and large stencil, respectively, and include both in our test cases since we expect them to have different performance characteristics.

Extensive literature exists on loop transformations and performance optimizations for stencil computations, which are often facilitated by the structured nature of stencil kernels. The common gather-like dataflow in stencil kernels, in which each iteration ``gathers'' data from a neighborhood to compute an update for a single point in the output, is trivial to parallelize, since each output is touched exclusively by one loop iteration, and the concurrent read access to overlapping neighborhoods is not problematic in a shared-memory environment. This dataflow pattern is replicated in the tangent-linear model, which thus inherits the original parallelism. In contrast, the dataflow reversal in the adjoint model results in a scatter-like data access in which each iteration reads from one point and ``scatters'' data to the neighborhood. Multiple iterations (and hence multiple threads) may attempt to update overlapping regions, and hence some safeguarding (in the form of atomics or reductions) is necessary. Previous work has observed this problem, and researchers have proposed loop transformations to implement adjoints of stencil loops as a gather operation~\cite{tfmad,adjstencil}. Outside of an AD context, a similar code transformation strategy has been proposed to reduce the load/store imbalance of primal stencil kernels~\cite{stock2014framework}. Using this strategy, a user can transform a gather-stencil into a scatter-stencil or even some intermediate shape where each iteration gathers and scatters from the same compact neighborhood. Performing these transformations is out of the scope of our work; but if this transformation was applied by the user to a primal stencil code, we can investigate the performance of AD applied to such a compact stencil. This is interesting because a side-effect of the compact stencil representation is that the read and write sets are identical for each iteration, and it has been observed in past work that shared-memory-parallel codes with identical read and write sets can be safely differentiated in reverse mode and retain their parallelism~\cite{ssmp} using the \texttt{shared} scope for the corresponding adjoint variables of all \texttt{shared} primal variables. We can therefore use the pragmas provided by Tapenade to enforce a \texttt{shared} scope, removing the need for reductions or atomics in this version. We therefore investigate the performance of two stencil variants: a \emph{conventional} variant that is implemented as a gather operation and a semantically equivalent \emph{compact} variant that is implemented by using identical read and write sets in every iteration.

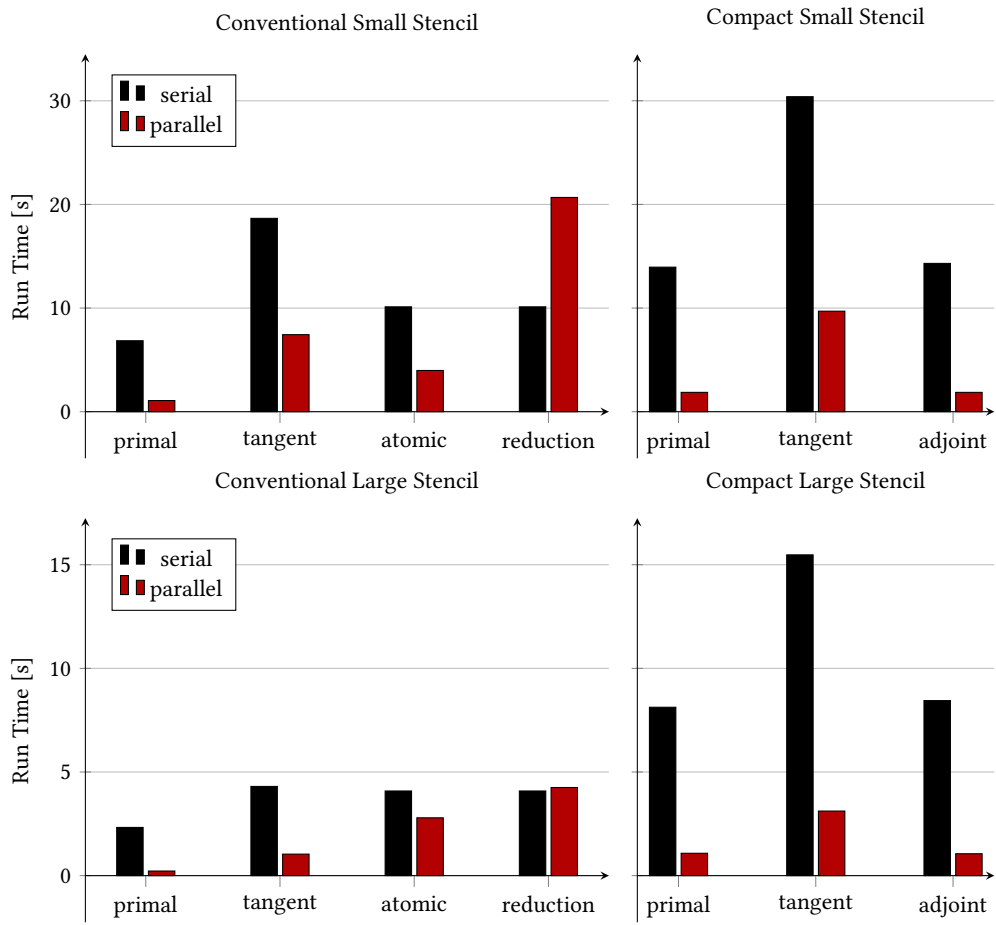
\begin{figure}
    \centering
\pgfplotstableread{
impl	time	serialtime
primal	1.076479912	6.847786188
tangent	7.434589863	18.65729189
atomic	3.978527069	10.12069917
reduction	20.6744101	10.12069917
}\timesdTable
\begin{tikzpicture}
    \begin{axis}
    [   axis y line=left,
        axis x line=middle,
        enlargelimits=0.15,
        ymin = 0,
        ymax = 30,
        ybar,
        ylabel={Run Time [s]},
        xticklabels from table={\timesdTable}{impl},
        xtick=data,
        ymajorgrids,
        title={Conventional Small Stencil},
        width=22em,
        height=17em,
        scale only axis,
        legend style={at={(0.05,0.95)}, anchor=north west}
    ]
        \addplot[fill=black] table [x expr=\coordindex, y=serialtime] {\timesdTable};\addlegendentry{serial};
        \addplot[fill=red!70!black] table [x expr=\coordindex, y=time] {\timesdTable};\addlegendentry{parallel};
    \end{axis}
\end{tikzpicture}
\pgfplotstableread{
impl	time	serialtime
primal	1.867021799	13.94784307
tangent	9.699846029	30.3970499
adjoint	1.866581917	14.31048608
}\timessTable
\begin{tikzpicture}
    \begin{axis}
    [   axis y line=left,
        axis x line=middle,
        enlargelimits=0.15,
        ymin = 0,
        ymax = 30,
        ybar,
        xticklabels from table={\timessTable}{impl},
        xtick=data,
        yticklabels={},
        ymajorgrids,
        title={Compact Small Stencil},
        width=15em,
        height=17em,
        scale only axis
    ]
        \addplot[fill=black] table [x expr=\coordindex, y=serialtime] {\timessTable};%\addlegendentry{serial};
        \addplot[fill=red!70!black] table [x expr=\coordindex, y=time] {\timessTable};%\addlegendentry{parallel};
    \end{axis}
\end{tikzpicture}
\pgfplotstableread{
impl	time	serialtime
primal	0.2223441601	2.32653904
tangent	1.038336992	4.304086924
atomic	2.790304899	4.084347486
reduction	4.252107859	4.084347486
}\timesdTable
\begin{tikzpicture}
    \begin{axis}
    [   axis y line=left,
        axis x line=middle,
        enlargelimits=0.15,
        ymin = 0,
        ymax = 15,
        ybar,
        ylabel={Run Time [s]},
        xticklabels from table={\timesdTable}{impl},
        xtick=data,
        ymajorgrids,
        title={Conventional Large Stencil},
        width=22em,
        height=17em,
        scale only axis,
        legend style={at={(0.05,0.95)}, anchor=north west}
    ]
        \addplot[fill=black] table [x expr=\coordindex, y=serialtime] {\timesdTable};\addlegendentry{serial};
        \addplot[fill=red!70!black] table [x expr=\coordindex, y=time] {\timesdTable};\addlegendentry{parallel};
    \end{axis}
\end{tikzpicture}
\pgfplotstableread{
impl	time	serialtime
primal	1.079764843	8.124480009
tangent	3.115815163	15.47664499
adjoint	1.056994915	8.444014072
}\timessTable
\begin{tikzpicture}
    \begin{axis}
    [   axis y line=left,
        axis x line=middle,
        enlargelimits=0.15,
        ymin = 0,
        ymax = 15,
        ybar,
        xticklabels from table={\timessTable}{impl},
        xtick=data,
        yticklabels={},
        ymajorgrids,
        title={Compact Large Stencil},
        width=15em,
        height=17em,
        scale only axis
    ]
        \addplot[fill=black] table [x expr=\coordindex, y=serialtime] {\timessTable};%\addlegendentry{serial};
        \addplot[fill=red!70!black] table [x expr=\coordindex, y=time] {\timessTable};%\addlegendentry{parallel};
    \end{axis}
\end{tikzpicture}
    \caption{\textbf{Top panels (left and right:)} Small stencil, absolute runtimes for 512 iterations on a 10,000,000-cell mesh. \textbf{Bottom panels (left and right:)} Large stencil, absolute runtimes for 64 iterations on a 10,000,000-cell mesh. \textbf{Left panels (top and bottom):} times for serial and parallel conventional stencil programs. There are two parallel adjoint variants, using either reductions or atomic updates. There is only one serial adjoint variant, compiled without OpenMP support; its timing is shown next to both parallel variants. \textbf{Right panels (top and bottom):} times for the compact stencil implementation. The compact stencil does not need atomics or reductions, and hence there is only one parallel variant that performs exactly as many operations and has the same dataflow as the primal, with identical runtimes in serial and parallel.
    }
    \label{fig:time_stencil_small}
\end{figure}

\pgfplotstableread{
threads	plaintime	compacttime	plaindtime	compactdtime	plainbrtime	plainbatime	compactbtime
1	1	1	1	1	0.4895278325	0.08534934236	1
2	1.94780034	1.874176273	1.492034874	1.572376388	0.4529579981	0.1673703146	1.975865208
4	3.875387347	3.760484348	2.070379351	2.421390569	0.2964131233	0.3176263464	3.929353502
8	5.304727115	5.036012943	2.35024308	2.841525905	0.2761138799	0.6269962707	5.191796356
16	6.361276335	6.416598362	2.509525372	3.133766228	0.1891618034	1.254209176	6.355324207
28	6.259813782	7.470637505	2.46100852	3.094031854	0.1249984426	1.97462102	7.666679908
56	6.203098213	7.375436038	2.342583384	3.008240885	0.06968267976	2.543830667	7.568182517
}\stencilsmalltable
\pgfplotstableread{
threads	plaintime	compacttime	plaindtime	compactdtime	plainbrtime	plainbatime	compactbtime
1	1	1	1	1	0.755084114	0.05899897619	1
2	1.905292353	1.876921534	1.564465906	1.777951348	0.9605465389	0.1163968844	1.960411565
4	3.320993717	3.806885398	2.416020348	3.167070712	0.8172972904	0.2204499757	4.014068204
8	6.988391129	5.056581294	3.446589332	4.207424602	0.8034036354	0.4345000991	5.324077931
16	9.300727801	6.378318963	4.122357597	4.967125514	0.595718697	0.8681897435	6.797678597
28	8.993602865	7.490829287	4.145173441	4.833456366	0.3975471259	1.373106297	7.98869886
56	10.46368404	7.524305002	3.904656653	4.806356285	0.225835761	1.463763866	7.928070914
}\stencillargetable
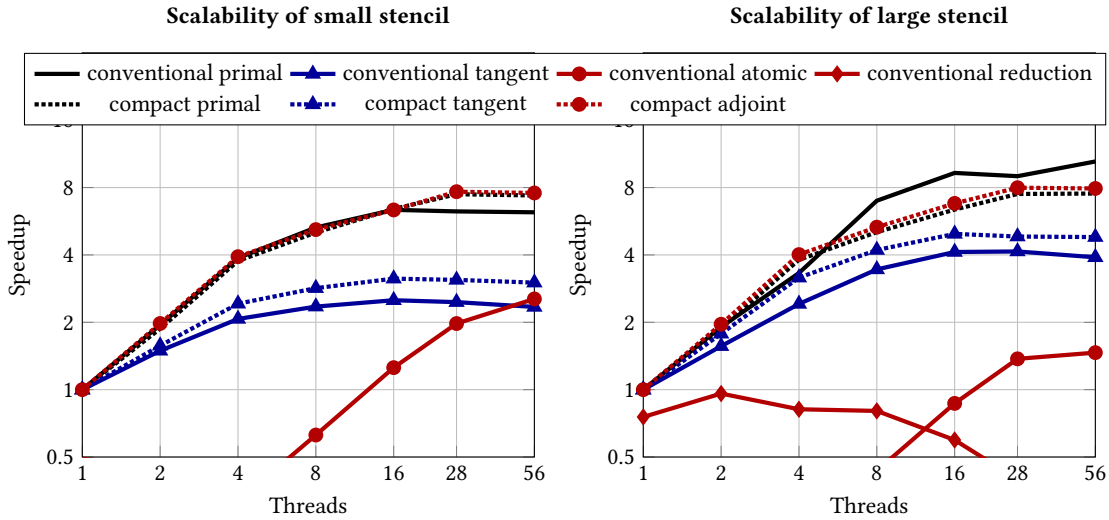
\begin{figure}[p]
\begin{tikzpicture}
\begin{axis}[
ymin=0.5,
ymax=32,
xmin=1,
xmax=56,
legend style={at={(0,1)},
      anchor=north west,legend columns=2},
xtick={1,2,4,8,16,28,56},
ytick={0.5,1,2,4,8,16,28},
xticklabels={1,2,4,8,16,28,56},
yticklabels={0.5,1,2,4,8,16,28},
xlabel={Threads},
ylabel={Speedup},
title={Scalability of small stencil},
title style={font=\bfseries},
xmode=log,
ymode=log,
xmajorgrids,
ymajorgrids,
width=19em,
height=17em,
cycle list name=cbw,
scale only axis
]
\addplot table [x expr=\thisrow{threads}, y expr=\thisrow{plaintime}] {\stencilsmalltable}; %\addlegendentry{plain primal}
\addplot table [x expr=\thisrow{threads}, y expr=\thisrow{plaindtime}] {\stencilsmalltable}; %\addlegendentry{plain tangent}
\addplot table [x expr=\thisrow{threads}, y expr=\thisrow{plainbatime}] {\stencilsmalltable}; %\addlegendentry{plain atomic}
\addplot table [x expr=\thisrow{threads}, y expr=\thisrow{plainbrtime}] {\stencilsmalltable}; %\addlegendentry{plain reduction}
\addplot table [x expr=\thisrow{threads}, y expr=\thisrow{compacttime}] {\stencilsmalltable}; %\addlegendentry{compact primal}
\addplot table [x expr=\thisrow{threads}, y expr=\thisrow{compactdtime}] {\stencilsmalltable}; %\addlegendentry{compact tangent}
\addplot table [x expr=\thisrow{threads}, y expr=\thisrow{compactbtime}] {\stencilsmalltable}; %\addlegendentry{compact adjoint}
\end{axis}
\end{tikzpicture}\hspace{-22.5em}
\begin{tikzpicture}
\begin{axis}[
ymin=0.5,
ymax=32,
xmin=1,
xmax=56,
legend style={at={(-1.37,1)},
      anchor=north west,legend columns=4},
xtick={1,2,4,8,16,28,56},
ytick={0.5,1,2,4,8,16,28},
xticklabels={1,2,4,8,16,28,56},
yticklabels={0.5,1,2,4,8,16,28},
xlabel={Threads},
ylabel={Speedup},
title={Scalability of large stencil},
title style={font=\bfseries},
xmode=log,
ymode=log,
xmajorgrids,
ymajorgrids,
width=19em,
height=17em,
cycle list name=cbw,
scale only axis
]
\addplot table [x expr=\thisrow{threads}, y expr=\thisrow{plaintime}] {\stencillargetable}; \addlegendentry{conventional primal}
\addplot table [x expr=\thisrow{threads}, y expr=\thisrow{plaindtime}] {\stencillargetable}; \addlegendentry{conventional tangent}
\addplot table [x expr=\thisrow{threads}, y expr=\thisrow{plainbatime}] {\stencillargetable}; \addlegendentry{conventional atomic}
\addplot table [x expr=\thisrow{threads}, y expr=\thisrow{plainbrtime}] {\stencillargetable}; \addlegendentry{conventional reduction}
\addplot table [x expr=\thisrow{threads}, y expr=\thisrow{compacttime}] {\stencillargetable}; \addlegendentry{compact primal}
\addplot table [x expr=\thisrow{threads}, y expr=\thisrow{compactdtime}] {\stencillargetable}; \addlegendentry{compact tangent}
\addplot table [x expr=\thisrow{threads}, y expr=\thisrow{compactbtime}] {\stencillargetable}; \addlegendentry{compact adjoint}
\end{axis}
\end{tikzpicture}
\caption{Scalability of stencil kernel. \textbf{Left:} Small stencil. The primal conventional and compact stencils scale almost perfectly up to 4 threads, then start to flatten out. The tangent programs scale less well, while the adjoint atomic scales very well up to 28 threads but starts from a very low point, since our scalability is plotted relative to the performance of the (much faster) serial program. Hence at least 16 threads are needed for the atomic parallel version to outperform the serial version. The reductions do not even appear in the plot, since they are too slow. The compact primal and adjoint have exactly the same performance profile. \textbf{Right:} Large stencil. Compared with the small stencil, scalability is overall slightly better for the primal and tangent, identical for the compact implementation, and slightly worse for the atomic adjoint. The worse performance of the atomic version can easily be explained by the need for even more atomic updates in each iteration. The reduction version does not outperform the serial program, and adding more threads increases runtime even further.}
\label{fig:scalestencillarge}
\end{figure}

In summary, this yields four stencil test cases: small conventional, small compact, large conventional, and large compact.
We show absolute runtimes in Figure~\ref{fig:time_stencil_small}. We can observe that the serial runtimes are best for the primal, slightly worse for the adjoint, and more than a factor of $2$ slower for the tangent than for the primal. The reason is that the tangent code contains both the primal result and the tangent derivative, while Tapenade by default removes any parts from the forward sweep that are not necessary to compute the adjoint. The serial runtimes of the compact stencil are worse than those of the conventional stencil, meaning that the gather-scatter transformation proposed in~\cite{stock2014framework} is not beneficial for serial performance in our test cases.

The primal parallel program operates on a one-dimensional mesh with 10 million cells. It uses a sequential outer loop that mimics 512 time steps for the small and 64 time steps for the large stencil. In each time step, an OpenMP worksharing loop is used to iterate over all mesh cells using a \texttt{static} schedule.

The parallel programs obtained from our new OpenMP-capable AD tool are significantly better than those of the serial programs in all cases except for the adjoint implementation using reductions. The reason is that the stencil case is already memory bandwidth bound; and the privatization, initialization, and combination of the privatized reduction arrays increase the overall amount of data transferred to and from memory by a factor that is almost identical to the number of threads. This is because for a domain of size $n$ and an execution on $T$ threads, each thread reads and writes only $\mathcal{O}(n/T)$ indices, so that all threads together access the entire array of size $n$. When using an OpenMP \texttt{reduction}, however, each thread will create and fully initialize a privatized version of the working arrays, leading to a total number of $\mathcal{O}(n\cdot T)$ memory accesses across all threads. For this reason, the runtime for the reduction variant actually \emph{increases} as more threads are added, as can be clearly seen in the scalability results in Figure~\ref{fig:scalestencillarge}. While the compact implementation was slower in serial, its adjoint is faster than any other variant in parallel, by more than a factor of $2$. The adjoint conventional stencil using \texttt{atomic} pragmas is faster than the serial adjoint, but the overhead of the atomic updates limits the overall parallel speedup particularly for the large stencil, which contains $17$ atomic increment operations in every iteration, compared with the small stencil containg $3$.

The scalability shown in Figure~\ref{fig:scalestencillarge} is less than ideal in either case. Since the processor used in this experiment has only 28 physical cores, we cannot expect good scalability beyond 28 threads. In fact, the scaling flattens out much earlier, and we achieve only around 8x speedup in either the small or large stencil case. The best scaling is achieved by the primal conventional and primal and adjoint compact implementations. The tangent programs scale more poorly, despite having the same parallel loop structure, scoping, and arithmetic intensity (that is, the ratio between floating-point operations and memory load/store operations). The reason is unclear.
Scalability of the large stencil is slightly better than that of the small stencil, probably because more floating-point operations are performed for every data point, increasing the arithmetic intensity and therefore making the code slightly less bandwidth bound, so that additional threads have a higher likelihood of helping performance.

We note that for the stencil test case, our initial experiments showed that the primal conventional stencil was significantly slower than the tangent or adjoint stencils. This was because the Intel compiler tried unsuccessfully to automatically vectorize the primal conventional stencil, resulting in a slowdown, while it did not attempt to do so for the tangent, adjoint, or compact programs. We explicitly disabled automatic vectorization in this test case using the \texttt{-no-vec} command line argument during compilation, which improved the conventional primal runtime and did not affect the other runtimes.

\subsection{Lattice-Boltzmann Solver}
In this test case, we use a simple implementation of the Lattice-Boltzmann Method~\cite{chen1998lattice} (LBM) solver from the Parboil benchmark suite~\cite{stratton2012parboil}. Since the Parboil benchmark is written in C and our current implementation only supports OpenMP in Fortran programs, we manually translated a core routine of the benchmark (less than 100 lines of code). LBM typically operates on a structured mesh and consists of a \emph{streaming} operation that involves communication between neighboring cells in the mesh, and a \emph{collision} operation that is compute intensive and local to each cell. LBM is thus somewhat similar to a stencil kernel but with a slightly higher arithmetic intensity.

The absolute runtime and scalability for this test case are shown in Figure~\ref{fig:scalelbm}. As expected, the tangent linear program is about twice slower than the primal program, and both scale equally well up to about 16 threads. Unlike in the stencil test case, the adjoint is significantly slower than the primal or tangent programs. The reason is that the collision step in LBM is a nonlinear operation, and Tapenade thus needs to implement a full forward sweep, store intermediate states in every forward iteration, and then retrieve those results in every reverse iteration, causing significant runtime overhead. Moreover, while LBM could also be implemented by using a compact dataflow just like the stencil kernel, the implementation in the Parboil benchmark suite has a gather-like dataflow that requires \texttt{atomic} pragmas or \texttt{reduction} clauses. The version using atomic updates is extremely slow in our experiment and never comes close to the performance of the serial adjoint program. The version using OpenMP reductions does outperform the serial program, albeit only by a factor of around 2 in the best case; and adding more than 8 threads leads to a slowdown, presumably because of the additional memory traffic.

\subsection{Green's Function Monte Carlo}
The Green's function Monte Carlo kernel\footnote{https://asc.llnl.gov/coral-benchmarks\#gfmcmk} is a method from nuclear physics and is part of the CORAL benchmark suite~\cite{vazhkudai2018design}. The primal program contains a loop in which each iteration can take a different amount of time and is thus parallelized by using an OpenMP \texttt{dynamic} schedule. The runtimes of the tangent and adjoint serial program are almost identical in this case, because the computation has only a few nonlinear operations and Tapenade thus inserts only a small amount of stack store and load operations, leading to an adjoint program that performs almost the same amount of work as the tangent program.

\begin{figure}
    \centering
\pgfplotstableread{
impl	serialtime	time
primal	13.12351489	0.7058608532
tangent	26.48179603	1.305618048
atomic	26.47304797	376.0379021
reduction	26.47304797	5.111660004
}\timesstassTable
\pgfplotstableread{
threads	plaintime		plaindtime		plainbatime	plainbrtime
1	1	1	0.07039994592	0.9585236227
2	1.963181789	2.191063657	0.2236093682	1.88208423
4	3.879220353	4.273120464	0.08101961584	3.448428399
8	7.404743318	8.026353069	nan	5.403052241
16	13.67922152	14.27878747	nan	5.072473115
28	18.59221238	18.05212202	nan	3.752934498
56	18.52413111	20.282958	nan	2.326132658
}\stassuijtable
\begin{tikzpicture}
    \begin{axis}
    [   axis y line=left,
        axis x line=middle,
        enlargelimits=0.15,
        ymin = 0,
        ymax = 34,
        ybar,
        xticklabels from table={\timesstassTable}{impl},
        xtick=data,
        ymajorgrids,
        title={Absolute Time},
        ylabel={Run Time [s]},
        width=19em,
        height=20em,
        scale only axis,
        title style={font=\bfseries},
        legend style={at={(0.05,0.95)}, anchor=north west}
    ]
        \addplot[fill=black] table [x expr=\coordindex, y=serialtime] {\timesstassTable};\addlegendentry{serial};
        \addplot[fill=red!70!black] table [x expr=\coordindex, y=time] {\timesstassTable};\addlegendentry{parallel};
        \coordinate(dl) at (2.0,37);% for the discontinuity
        \coordinate(dr) at (2.3,37);% for the discontinuity
        \coordinate(ul) at (2.0,40);% for the discontinuity
        \coordinate(ur) at (2.3,40);% for the discontinuity
    \end{axis}
  \filldraw[fill=white] (dl)
  [snake=zigzag]     -- (dr)
  [snake=none,white] -- (ur)
                     -- (ul);
\end{tikzpicture}
\begin{tikzpicture}
\begin{axis}[
ymin=0.5,
ymax=32,
xmin=1,
xmax=56,
legend style={at={(0,1)},
      anchor=north west,legend columns=1},
xtick={1,2,4,8,16,28,56},
ytick={0.5,1,2,4,8,16,28},
xticklabels={1,2,4,8,16,28,56},
yticklabels={0.5,1,2,4,8,16,28},
xlabel={Threads},
ylabel={Speedup},
title={Scalability},
title style={font=\bfseries},
xmode=log,
ymode=log,
xmajorgrids,
ymajorgrids,
width=19em,
height=17em,
cycle list name=cbw,
scale only axis
]
\addplot table [x expr=\thisrow{threads}, y expr=\thisrow{plaintime}] {\stassuijtable}; \addlegendentry{primal}
\addplot table [x expr=\thisrow{threads}, y expr=\thisrow{plaindtime}] {\stassuijtable}; \addlegendentry{tangent}
\addplot table [x expr=\thisrow{threads}, y expr=\thisrow{plainbatime}] {\stassuijtable}; \addlegendentry{atomic}
\addplot table [x expr=\thisrow{threads}, y expr=\thisrow{plainbrtime}] {\stassuijtable}; \addlegendentry{reduction}
\end{axis}
\end{tikzpicture}
\caption{\textbf{Left:} Absolute runtimes for GFMC test case. The atomic parallel time is 376 s and is truncated in the plot. \textbf{Right:} Scalability for primal and tangent GFMC is equally good but is significantly worse for adjoint reductions. The adjoint program using atomics scales well but performs too poorly compared with the serial adjoint program to show up in the plot.}
\label{fig:scalestassuij}
\end{figure}
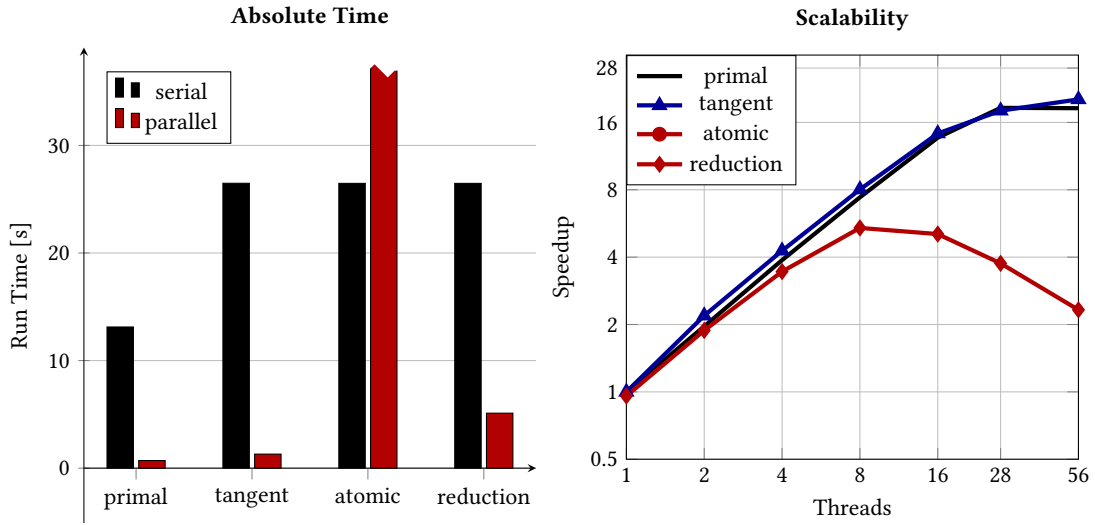

Just as in the LBM case, atomic updates lead to a performance overhead that leads to worse performance than in the serial case. The program containing reductions scales reasonably well up to 8 threads, then slows as more threads are added. The runtimes and scalability are shown in Figure~\ref{fig:scalestassuij}.

\begin{figure}
    \centering
\pgfplotstableread{
impl	serialtime	time
primal	1.691883	0.208471
tangent	3.914721	0.459241
atomic	10.691821	33.208728
reduction	10.691821	4.998249
}\timeslbmTable
\pgfplotstableread{
threads	plaintime		plaindtime		plainbatime	plainbrtime
1	1.003558166	1.003286313	0.3219581611	0.8971583552
2	1.885308301	1.873164613	0.1463825624	1.42450054
4	3.525136197	3.50394503	0.05995460084	1.723811079
8	6.588706204	6.371162584	0.05903315936	2.139113317
16	7.959753599	8.481002108	0.05815984986	1.981800384
28	8.14455248	8.552341799	0.05699019193	1.481096517
56	8.119159538	8.344280336	0.05515796338	0.9004324203
}\lbmtable
\begin{tikzpicture}
    \begin{axis}
    [   axis y line=left,
        axis x line=middle,
        enlargelimits=0.15,
        ymin = 0,
        ymax = 15,
        ybar,
        xticklabels from table={\timeslbmTable}{impl},
        xtick=data,
        ymajorgrids,
        title={Absolute Time},
        title style={font=\bfseries},
        ylabel={Run Time [s]},
        width=19em,
        height=20em,
        scale only axis,
        legend style={at={(0.05,0.95)}, anchor=north west}
    ]
        \addplot[fill=black] table [x expr=\coordindex, y=serialtime] {\timeslbmTable};\addlegendentry{serial};
        \addplot[fill=red!70!black] table [x expr=\coordindex, y=time] {\timeslbmTable};\addlegendentry{parallel};
        \coordinate(dl) at (2.0,16);% for the discontinuity
        \coordinate(dr) at (2.3,16);% for the discontinuity
        \coordinate(ul) at (2.0,17.5);% for the discontinuity
        \coordinate(ur) at (2.3,17.5);% for the discontinuity
    \end{axis}
  \filldraw[fill=white] (dl)
  [snake=zigzag]     -- (dr)
  [snake=none,white] -- (ur)
                     -- (ul);
\end{tikzpicture}
\begin{tikzpicture}
\begin{axis}[
ymin=0.5,
ymax=16,
xmin=1,
xmax=56,
legend style={at={(0,1)},
      anchor=north west,legend columns=1},
xtick={1,2,4,8,16,28,56},
ytick={0.5,1,2,4,8,16,28},
xticklabels={1,2,4,8,16,28,56},
yticklabels={0.5,1,2,4,8,16,28},
xlabel={Threads},
ylabel={Speedup},
title={Scalability},
title style={font=\bfseries},
xmode=log,
ymode=log,
xmajorgrids,
ymajorgrids,
width=19em,
height=17em,
cycle list name=cbw,
scale only axis
]
\addplot table [x expr=\thisrow{threads}, y expr=\thisrow{plaintime}] {\lbmtable}; \addlegendentry{primal}
\addplot table [x expr=\thisrow{threads}, y expr=\thisrow{plaindtime}] {\lbmtable}; \addlegendentry{tangent}
\addplot table [x expr=\thisrow{threads}, y expr=\thisrow{plainbatime}] {\lbmtable}; \addlegendentry{atomic}
\addplot table [x expr=\thisrow{threads}, y expr=\thisrow{plainbrtime}] {\lbmtable}; \addlegendentry{reduction}
\end{axis}
\end{tikzpicture}
\caption{\textbf{Left:} Absolute runtimes for LBM test case. The atomic parallel time is 33 s and is truncated in the plot. \textbf{Right:} Scalability for primal and tangent LBM is equally good (the system has only 28 physical cores, hence poor scalability can be expected beyond that), but is significantly worse for adjoint reductions. The adjoint program using atomics scales well but performs too poorly compared with the serial adjoint program to show up in the plot.}
\label{fig:scalelbm}
\end{figure}
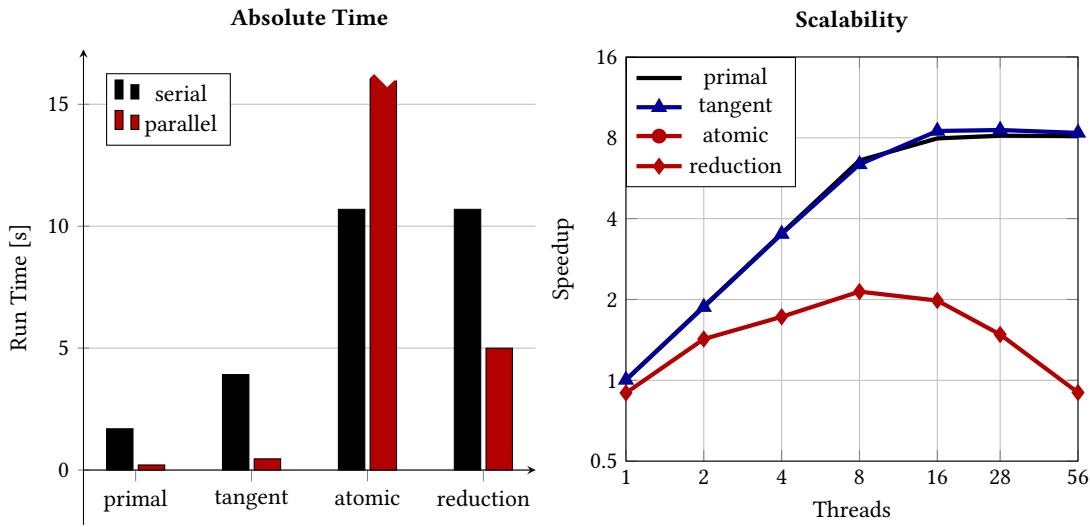

\section{Related Work}
\label{sec:related}

With automatic differentiation being applied to increasingly large applications, issues related to high-performance computing are often explored. It is important to preserve or retrieve the parallel qualities of the primal code in its tangent and adjoint derivatives. Several articles report on parallel adaption of differentiated codes, most often as a postprocessing stage after differentiation, for example for codes where the primal and adjoint can be parallelized in the same way because of read and write sets being identical, as in the case of the compact stencil in our paper~\cite{ssmp}. Other works have focused exclusively on adjoint differentiation of stencil-like operations~\cite{tfmad,adjstencil} but are restricted to that one type of input program and have not been implemented in a general-purpose AD tool. Yet other work discusses adding OpenMP pragmas into differentiated code in a post-processing step, OpenMP-parallelization of derivatives of sequential programs~\cite{Bucker2001BTA,Bucker2002ELS}, or AD applied to OpenMP parallel programs that have been modified to hide their parallelism from the AD tool~\cite{Wolf:64281,ssmp}.

Studies of full or partial preservation of parallel properties by the AD tool itself have focused mostly on coarse-grained parallel dialects such as MPI~\cite{Giering2005TLa, Ozkaya2012AoA, Towara2018Sam, Huck2018Auc, Cardesa2020Acb}.
MPI-like parallelism uses distributed memory, thus hiding the race conditions problems that we face with OpenMP. Other studies targeting GPU architectures~\cite{Grabner2008ADf,revels2018dynamic} share more similarities with OpenMP. In general, good performance is achieved, although help is often required from the end user (pragmas or postprocessing) because of limitations of static dataflow analysis.

A few studies address OpenMP specifically. In~\cite{Bischof2008PRM, Letschert2012ESi}, the AD tool is ADOL-C, belonging to the so-called \emph{operator overloading} class, for which parallel-preserving AD is made technically harder by the limited ability of the tools to analyze and transform the source and by the mixed nature of the differentiated code, with an initial phase in the application language and a second phase, often in C, that interprets the tape. Still, the essential parallelism questions are the same. For tools of the source transformation class, Giering et al.~\cite{Giering2005TLa} present adjoint AD of a large application from Earth sciences that combines MPI and OpenMP parallel models but provides little detail on the AD strategies on parallel constructs. Some such strategies are proposed in~\cite{Giering1996RfA}. In~\cite{forster2014algorithmic} the author presents a theoretical model for the differentiation of pragma-defined parallel regions, but the work is restricted to a simplified language called \emph{SPL} that has been extended with selected OpenMP pragmas.

In its so-called vector mode, AD produces a differentiated code that propagates derivatives along several directions in a single run. This can compute Jacobian matrices efficiently. In~\cite{Bucker2004ACo}, the authors observe that this introduces an extra level of iteration, which is obviously parallel and amenable to OpenMP. The question is then how to combine this new level of parallelism to the original OpenMP level that may come from the primal code.

Regarding the issue of ensuring atomicity of adjoint increments of shared variables in OpenMP, a few studies~\cite{Giering2005TLa, Bischof2008PRM} have identified the two alternatives---reduction or {\tt atomic} clause---but reach no consensus on the choice criteria. We hope that the differentiation model that we propose and our runtime experiments might clarify this choice.
There is also recent work presenting an efficient solution in an operator-overloading setting~\cite{Kaler2021PAW}; investigating whether this strategy can be adapted to source-transformation AD  would be interesting for future work.

Few studies provide a framework to validate their decisions, or their AD tool's decision, about the differentiated OpenMP directives that they introduce.
Although these decisions seem reasonable and final execution of the differentiated code is correct, we believe a justification framework such as the one in~\cite{Naumann2008AFf} is useful to gain confidence in these decisions or even to improve them.

Most studies mainly address the parallel dialect, for example the sublanguage of OpenMP, that is effectively used in the targeted final application code. In contrast, we selected a widely used subset of the OpenMP features, and we studied all the possible scoping pragmas that may occur in a worksharing loop, in combination with all possible worksharing schedules.

Apart from a mention in~\cite{Bischof2008PRM}, no study details the needed developments for adapting the AD runtime support to OpenMP. In particular adjoint AD, in its source-transformation as well as operator-overloading fashions, must store large amounts of data in the stack or tape. Using the {\tt threadprivate} declaration is generally seen as the right way to do so in an OpenMP environment, but we believe an in-depth description like the one we provide was needed.

\section{Conclusion, Future Work}
\label{sec:conclusion}
We have presented a model for the tangent-linear and adjoint differentiation of programs containing OpenMP parallel worksharing loops, and we have implemented that model in the Tapenade automatic differentiation tool. We have evaluated the performance of generated derivative codes on a set of scientific computing test cases.

Our derivative codes have achieved a speedup through parallelization compared with their serial counterparts in all our test cases. The scalability of the derivative codes, however, was worse than that of the original codes in some cases. Additionally, there is a choice between atomic updates (causing a run time overhead) and reductions (causing a memory and run time overhead) for the adjoint, where the best choice was case-dependent. An interesting task for future work would be to develop ways to automatically choose the best strategy, or at least to investigate properties that make a given program more likely to benefit from either approach. In the meantime, users can use our new pragmas to override the choice that Tapenade makes, if necessary.

Since neither atomic updates nor reductions offered satisfactory scalability for the adjoints, important directions for future work include developing new kinds of data-dependence analysis to automatically detect the absence of conflicting updates in the adjoint, in order to achieve the kinds of speedup that we obtained in the compact stencil. Since our adjoint test cases (and possibly many other real-world applications) contain parallel updates to the same output array with a low rate of conflicts between threads, it might also be profitable to develop a solution that can perform this operation more efficiently than either OpenMP atomics or reductions.

Other future work should include support (both in theory and in implementation) for a larger subset of OpenMP, including other (non-linear) reduction operators, SIMD pragmas, critical sections, tasks, and device offloading. A theory for correct differentiation of such codes could also be suitable for developing differentiation capabilities for other shared-memory parallelism languages such as OpenACC, CUDA, Pthreads, or OpenCL.

\begin{acks}
This work was funded in part by support from the U.S. Department of Energy, Office of Science, under contract DE-AC02-06CH11357. We gratefully acknowledge the computing resources provided and operated by the Joint Laboratory for System Evaluation (JLSE) at Argonne National Laboratory. We thank Johannes Doerfert for the fruitful discussions about OpenMP dynamic schedules that informed our schedule recording implementation.
\end{acks}

\bibliographystyle{ACM-Reference-Format}
\bibliography{main}

\end{document}